\documentclass[aps,pre,twocolumn,eqsecnum]{revtex4}
\usepackage{etoolbox}
\csundef{equation*}
\csundef{endequation*}
 
\usepackage{graphicx}
\usepackage{amsmath}
\usepackage{amssymb}
\usepackage{epstopdf}
\usepackage{epsfig}
\usepackage{color}
\usepackage{hyperref}
\usepackage{subfigure}

\def\be{\begin{equation}}
\def\ee{\end{equation}}
\def\bea{\begin{eqnarray}}
\def\eea{\end{eqnarray}}

\newcommand{\compC}{\mathbb{C}}
\let\Im\relax
\let\Re\relax
\DeclareMathOperator{\Im}{Im}
\DeclareMathOperator{\Re}{Re}
\DeclareMathOperator{\eff}{eff}

\begin{document}
\title{Nonmonotonic confining potential and eigenvalue density transition for generalized random matrix model}
\author{Swapnil Yadav$^1$, Kazi Alam$^1$, K. A. Muttalib$^1$ and Dong Wang$^2$}
\address
{$^1$Department of Physics, University of Florida, Gainesville, Florida 32611-8440, USA\\
$^2$Department of Mathematics, National University of Singapore, Singapore 119076}
\date{Apr 27, 2021}

\begin{abstract}
We consider several limiting cases of the joint probability distribution for a random matrix ensemble with an additional interaction term controlled by an exponent $\gamma$ (called the $\gamma$-ensembles). The effective potential, which is essentially the single-particle confining potential for an equivalent ensemble with $\gamma=1$ (called the Muttalib-Borodin ensemble), is a crucial quantity defined in solution to the Riemann-Hilbert problem associated with the $\gamma$-ensembles. It enables us to numerically compute the eigenvalue density of $\gamma$-ensembles for all $\gamma > 0$. We show that one important effect of the two-particle interaction parameter $\gamma$ is to generate or enhance the non-monotonicity in the effective single-particle potential. For suitable choices of the initial single-particle potentials, reducing $\gamma$ can lead to a large non-monotonicity in the effective potential, which in turn leads to significant changes in the density of eigenvalues. For a disordered conductor,  this corresponds to a systematic decrease in the conductance with increasing disorder. This suggests that appropriate models of $\gamma$-ensembles can be used as a possible framework to study the effects of disorder on the distribution of conductances.

\end{abstract} 

\maketitle

\section{Introduction}
A generalized random matrix model with additional interactions \cite{Alam-Muttalib-Wang-Yadav20}, called the $\gamma$-ensembles, was introduced recently as a solvable toy model for three-dimensional (3D) disordered conductors. The joint probability distribution (jpd) of the $N$ non-negative eigenvalues $x_i$ for these $\gamma$-ensembles has the form
\begin{equation}
\begin{aligned}
&p(\{x_i\};\theta,\gamma) \propto \prod_{i=1}^Nw(x_i)\prod_{i<j}|x_i-x_j||x_i^{\theta}-x_j^{\theta}|^{\gamma},\\ 
&0< \gamma, \;\;\;  1 < \theta < \infty.
\label{gamma_ensemble_jpd}
\end{aligned}
\end{equation}
Here we assume the convention $w(x) = e^{-NV(x)}$, so that the empirical distribution of the particles (a.k.a.  the equilibrium measure) converges as $N \to \infty$.  In \cite{Alam-Muttalib-Wang-Yadav20}, the parameter $\gamma$ was restricted to $ 0 < \gamma \le 1$, but the method developed there allows the evaluation of the density of eigenvalues of the $\gamma$-ensembles for any $\gamma > 0$, $\theta>1$ and for any well behaved $V(x)$. In particular, it was shown that the jpd for the $\gamma$-ensembles can be mapped on to the  Muttalib-Borodin (MB) ensembles \cite{Muttalib95,Borodin99,Forrester-Wang15,Zhang15,Kuijlaars-Molag19,Molag20,Wang-Zhang21} (which has the same jpd as Eq. (\ref{gamma_ensemble_jpd}), with $\gamma=1$), by replacing the external potential $V(x)$ with a $\gamma$-dependent  effective potential $V_{\eff}(x;\gamma)$. This effective potential was calculated explicitly for $\theta=2$ by numerically solving the Riemann-Hilbert (RH) problem associated with the jpd of the $\gamma$-ensembles. This allowed the calculation of the corresponding exact density of the eigenvalues $\sigma(x)$, which can be used to calculate the average conductance of a disordered conductor. 

In terms of the variables in Eq. (\ref{gamma_ensemble_jpd}), the average dimensionless conductance per channel $g_{channel}$ of a disordered conductor (in units of the quantum conductance $e^2/\hbar$) is given by \cite{note,Muttalib-Pichard-Stone87}
\be
g_{channel}=\int_0^{\infty} \frac{\sigma(x)}{\cosh^2\sqrt{x}}dx.
\label{g}
\ee
Clearly, a large peak in the density near the origin corresponds to a large conductance, or a metal, while a density which is small near the origin and spread out at large values of $x$ will correspond to a small conductance, or an insulator.

As shown in \cite{Alam-Muttalib-Wang-Yadav20}, while the exact solution of the density for Eq. (\ref{gamma_ensemble_jpd}) for $\theta=2$ shows a significant change in the density as a function of the two-particle interaction parameter $\gamma$, the change in density is not large enough to affect the conductance $g$ significantly. Thus the question arises: What is the role of the parameter $\gamma$ in the transition from metallic to insulating behavior of a disordered quantum conductor?
In this paper we address this question in three steps: 

First, we show that if we allow $1 < \theta < 2$, then the effective potential near the origin becomes non-monotonic for $\gamma < 1$, where the degree of non-monotonicity increases with decreasing $\gamma$. This is significant because such non-monotonic effective potential can in principle give rise to a transition in density from hard-edge to soft-edge, which means a transition from a diverging to a non-diverging density near the origin, as shown by Clays and Romano (CR) \cite{Claeys-Romano14}. As a bonus, we find that for Laguerre $\beta$ ensembles, the eigenvalue density for all values of $\beta\ge 1$ can be obtained by considering the $\theta\to 1$ limit of the $\gamma$-ensembles, with $\beta=\gamma+1$, as shown in Appendix A. 

Second, while the CR model (which belongs to the MB-ensembles) shows a transition from a diverging to a non-diverging density near the origin by changing the non-monotonicity parameter $\rho$ of the single-particle potential $V(x)=x^2-\rho x$, we show that for a fixed value of $\rho$, a similar transition occurs as a function of the two-particle interaction parameter $\gamma$. This shows that the role of the parameter $\gamma$ in the $\gamma$-ensembles is qualitatively similar to a non-monotonicity  parameter in the single-particle potential. 

Third, we consider a realistic phenomenological single-particle potential for a disordered conductor of the form $V(x)=\Gamma x - (1/2)\ln \sinh^2\sqrt{x}$ where the logarithmic term arises naturally as a Jacobian factor \cite{Markos-Muttalib-Wolfle05} and $\Gamma$ is also a function of $\gamma$. This model produces a transition in the density from a peak near the origin to a density with a gap near the origin as $\gamma$ is reduced systematically from 1, the gap increasing with decreasing $\gamma$. This change in the density is sufficient to result in a transition from a metallic to an insulating conductance.  
While such a toy model is clearly not sufficient to describe metal-to-insulator transition in actual physical systems,  the results suggest that the $\gamma$-ensembles with appropriate single-particle potentials can be used as a possible framework to study the distribution of conductances across the metal-insulator transition.

The paper is organized as follows. In Sec. \ref{sec:2} we give a brief outline of the numerical solution to RH problem for $\gamma$ ensembles. The equilibrium density can be obtained replacing external potential $V(x)$ with $\gamma$ dependent effective potential $V_{\eff}(x;\gamma)$. In Secs. \ref{sec:3} , \ref{sec:4}  and \ref{sec:5} we follow the three steps  mentioned above and systematically explore the role of the parameter $\gamma$. 
We summarize our results  in Sec. \ref{sec:6}.  Results obtained as a bonus for the well-known 
$\beta$-ensembles as a $\theta \to 1$ limit of the $\gamma$-ensembles are discussed in Appendix A. Some mathematical details are given in Appendix B.

 
\section{The equilibrium problem for $\gamma$ ensemble} \label{sec:2}

\begin{figure}
\begin{center}
\includegraphics[width=0.5\textwidth]{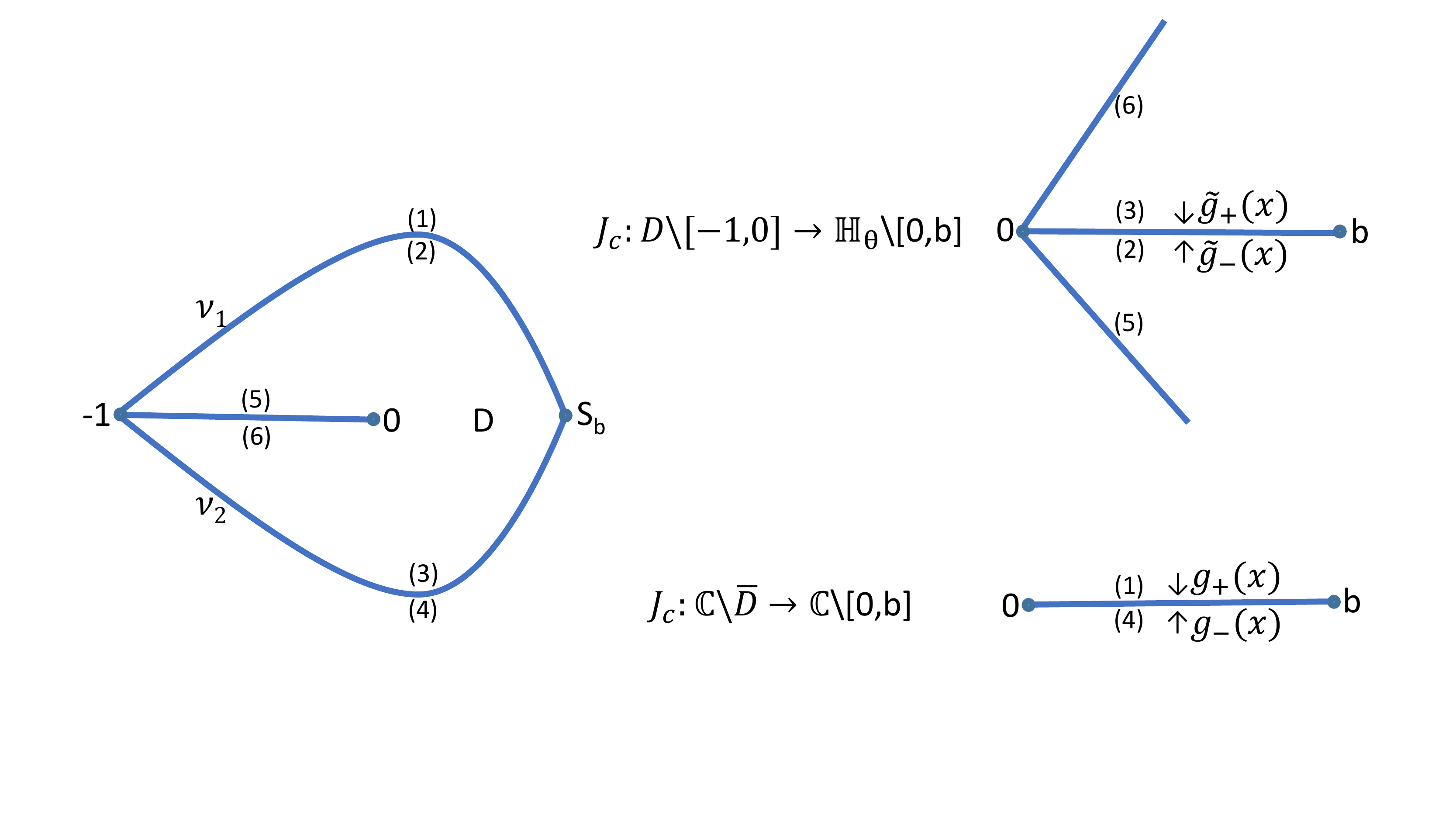}
\end{center}
\caption{
(Color online) 
Schematic figure for the mapping of JT for a hard-edge problem. Here $D$ is the region inside the contour $\nu_1,\nu_2$ ($\bar{D}$ is the region outside). $\mathbb{H}_\theta$ is the angular region at the top right between the lines [5],[6]. $\mathbb{C}$ denotes the complex plane.}
\label{mapping_schematic}
\end{figure}

Here we give a brief overview of the solution to the RH problem of $\gamma$-ensembles and the computation of its eigenvalue density. The complete analysis can be found in \cite{Alam-Muttalib-Wang-Yadav20}. Consider the $\gamma$-ensembles defined by the jpd in Eq. (\ref{gamma_ensemble_jpd}).
The unique equilibrium measure $\mu$ that minimizes the energy functional 
\begin{equation}
\begin{aligned}
  &\frac{1}{2} \iint \ln \frac{1}{\lvert x - y \rvert} d\mu(x) d\mu(y) + \frac{\gamma}{2} \iint \ln \frac{1}{\lvert x^{\theta} - y^{\theta} \rvert} d\mu(x) d\mu(y) \\ 
  &+ \int V(x) d\mu(x),
\end{aligned}  
\end{equation}
satisfies the Euler-Lagrange (EL) equation
\begin{equation}
\int \ln| x - y| d\mu(y) +\gamma\int \ln| x^{\theta}-y^{\theta}|d\mu(y) 
- V(x)= \ell
\label{euler-lagrange}
\end{equation}
if $x$ lies inside the support of density and the equality sign is replaced by  $<$ if $x$ lies outside the support. Here $\ell$ is some constant. In this section we give formalism for hard-edge support where we assume that the eigenvalue density lies on support $[0,b]$ for some $b>0$. The similar formalism for soft-edge for which density lies on support $[a,b]$ with $b>a>0$, is given in Appendix B. In formulating the RH problem from the above EL equations, crucial role is played by the Joukowsky transformation (JT) for hard edge,    
\begin{equation}
\begin{aligned}
  J_c(s) = {}& c(s+1)(\frac{s+1}{s})^{\frac{1}{\theta}}, \\
\end{aligned}
\label{joukowsky}
\end{equation}  
where $s$ is a complex variable. The points in the complex domain, which are mapped by the JT on to a real line, form a contour $\nu$ given by,
\begin{equation}
r(\phi)= \left. \tan \left( \frac{\phi}{1+\theta} \right) \middle/ \left[ \sin\phi-\cos\phi\tan \left( \frac{\phi}{1+\theta} \right) \right] \right.,
\label{contour_hard_edge}
\end{equation}
where $0<\phi<2\pi$ is the argument of $s$ in the complex plane. Schematic Fig. \ref{mapping_schematic} shows mapping of all points on contour $\nu$ to two different regions in the complex plane by the JT $J_c(s)$.
By defining complex transforms
\begin{equation}
\begin{aligned}
  g(z) \equiv {}& \int_0^b \log(z-x)d\mu(x), && z \in \mathbb{C}\backslash(-\infty,b], \\
  \tilde{g}(z) \equiv {}& \int_0^b \log(z^{\theta}-x^{\theta})d\mu(x), && z \in \mathbb{H}_\theta\backslash(0,b],
\end{aligned}
\label{complex_transforms}
\end{equation}
with their derivatives $G(s)\equiv g^{\prime}(s)$, $\tilde{G}(s)\equiv \tilde{g}^{\prime}(s)$ and the function $M(s)$ as,
\begin{equation}
M(s)\equiv
\begin{cases}
  G[J_c(s)], & \text{for } s\in\mathbb{C}\backslash \bar{D}, \\
  \tilde{G}[J_c(s)], & \text{for } s\in D\backslash[-1,0],
\end{cases}
\label{M_def}    
\end{equation}
the sum and difference of the EL equations can be written as
\begin{equation}
\begin{split}
  M_+(s_1)+\gamma M_-(s_1)+M_-(s_2)+\gamma M_+(s_2) 
  = {}& 2V^{\prime}[J_c(s)], \\
  M_+(s_1)-M_-(s_2)+M_-(s_1)-M_+(s_2) 
  = {}& 0.
\end{split}
\label{M_EL}
\end{equation}
Here $s_1 \in \nu_1$ and $s_2 \in \nu_2$ (see Fig. \ref{mapping_schematic}). Equation (\ref{M_EL}), together with some of the limits of $M(s)$, form the RH problem for $M(s)$. The RH problem in terms of $N(s)\equiv M(s)J_c(s)$ is then as follows.

\vspace{0.2cm}

\paragraph*{RH problem for $N$:}

\begin{itemize}
\item
  $N$ is analytic in $\compC \setminus \nu$.
\item 
      $N_+(s_1)+\gamma N_-(s_1)+N_-(s_2)+\gamma N_+(s_2)$\\
      $=  2V^{\prime}[J_c(s)]J_c(s)$ 
      \begin{equation}
      N_+(s_1)-N_-(s_2)+N_-(s_1)-N_+(s_2) 
      =  0.
    \label{Npm}
  \end{equation}
\item 
  $N(0) = \theta$ and $N(s) \to 1$ as $s \to \infty$.
\end{itemize}
We further define a function $f$ such that,
\begin{equation} \label{eq:defn_f}
f[J_c(s)]\equiv N_+(s)+N_-(s).
\end{equation}
This gives solution to RH problem of $N(s)$ as,
\begin{equation}
N(s)=
\begin{cases}
  \frac{-1}{2\pi i}\oint_{\nu}\frac{f[J_c(\xi)]}{\xi -s}\; d\xi +1, & s\in \mathbb{C}\backslash \bar{D}, \cr
\frac{1}{2\pi i}\oint_{\nu}\frac{f[J_c(\xi)]}{\xi -s}\; d\xi -1, & s\in D\backslash [-1,0].
\end{cases}
\label{N_def_contr}
\end{equation}
Also from the RH problem for $N(s)$, the constant $c$ of the JT in Eq. (\ref{joukowsky})  satisfies the equation
\begin{equation}
\label{c_hard_edge}
\frac{1}{2\pi i}{\displaystyle \oint_{\nu}^{}}\frac{f[J_c(s)]}{s}ds=1+\theta.
\end{equation}
Thus the sum equation in the RH problem for $N(s)$ can be rewritten as,
\begin{equation}
(1-\gamma)[N_+(s_1)+N_-(s_2)]+2\gamma f[J_c(s)] 
=2V^{\prime}[J_c(s)]J_c(s). 
\end{equation}
Defining the inverse mapping of JT as,
\begin{equation}
s=J_c^{-1}(x)=h(x).
\label{inversemap}
\end{equation}
with $(s_1)_+ = h(y); \ (s_2)_- = \bar{h}(y); \ s_1 = h(x) \ \text{and} \ s_2 = \bar{h}(x) $, we substitute for $[N_+(s_1)+N_-(s_2)]$ using Eq. (\ref{N_def_contr}) and the inverse mapping. We finally get the integral equation,
\begin{equation}
\label{f_integral_eqn}
  f(y;\gamma) = \frac{V^{\prime}(y)y}{\gamma} 
  -\frac{1-\gamma}{\gamma}\bigg[1 + \frac{1}{2\pi}\int_0^b f(x;\gamma)\phi(x,y)dx \bigg],
\end{equation}
where
\begin{equation}
  \phi(x,y) = \Im\bigg[ \left( \frac{1}{h(y) - \overline{h}(x)} + \frac{1}{\overline{h}(y) - \overline{h}(x)} \right) \overline{h}^{\prime}(x) \bigg].
\end{equation} 
We solve Eq. (\ref{f_integral_eqn}) for $f(y;\gamma)$ and Eq. (\ref{c_hard_edge}) for $c$ numerically, self-consistently. The new effective potential $V_{\eff}(x;\gamma)$ is related to $f(x;\gamma)$ by
\begin{equation}
V'_{\eff}(x;\gamma)=\frac{f(x;\gamma)}{x}.
\label{V-effective}
\end{equation}
 The eigenvalue density for this effective potential is given by \cite{Alam-Muttalib-Wang-Yadav20},  
\begin{equation}
\label{density_hard_edge}
\begin{split}
  \sigma(y;\gamma)= {}& \frac{-1}{2{\pi}^2 \gamma y}\int_{b}^0 xV'_{\eff}(x;\gamma)\chi(x,y) dx,\\
  \chi(x,y)= {}& \Re\bigg[ \bigg( \frac{1}{\overline{h}(y) - h(x)}-\frac{1}{h(y) - h(x)} \bigg)h^{\prime}(x)\bigg].
\end{split}
\end{equation}

\begin{figure*}
    \centering
    \subfigure{\includegraphics[width=0.325\textwidth]{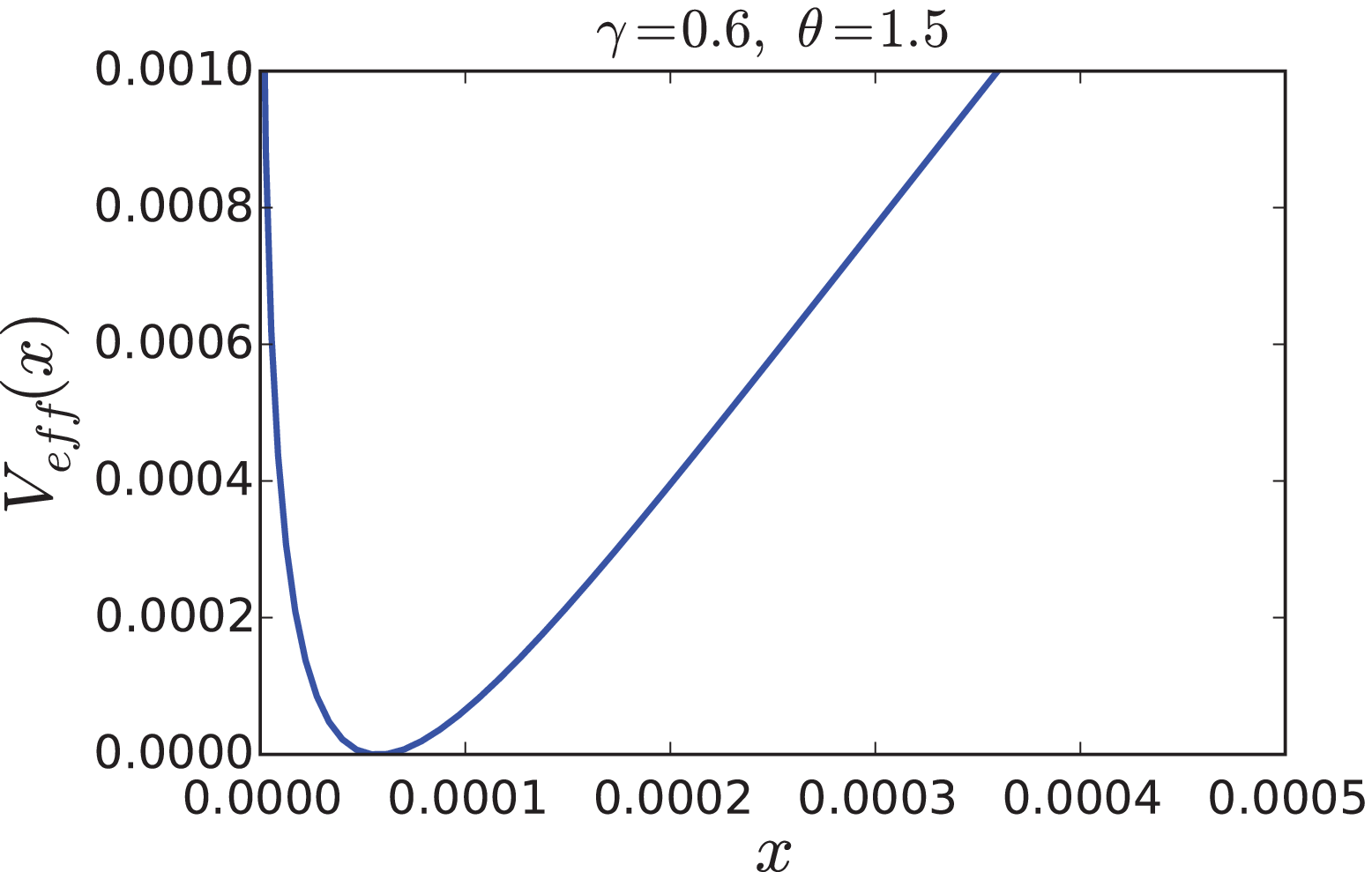}} 
    \subfigure{\includegraphics[width=0.325\textwidth]{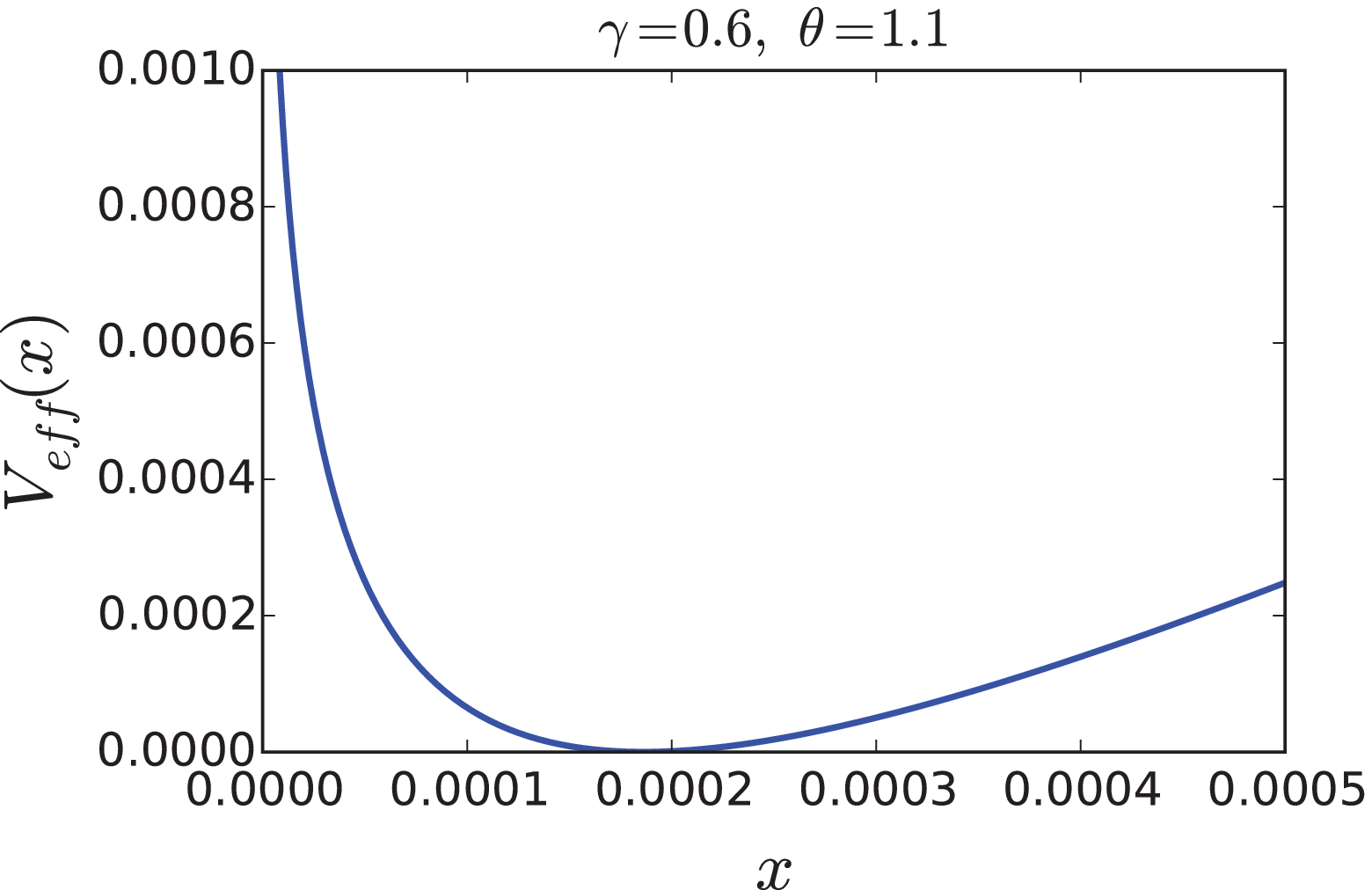}} 
    \subfigure{\includegraphics[width=0.325\textwidth]{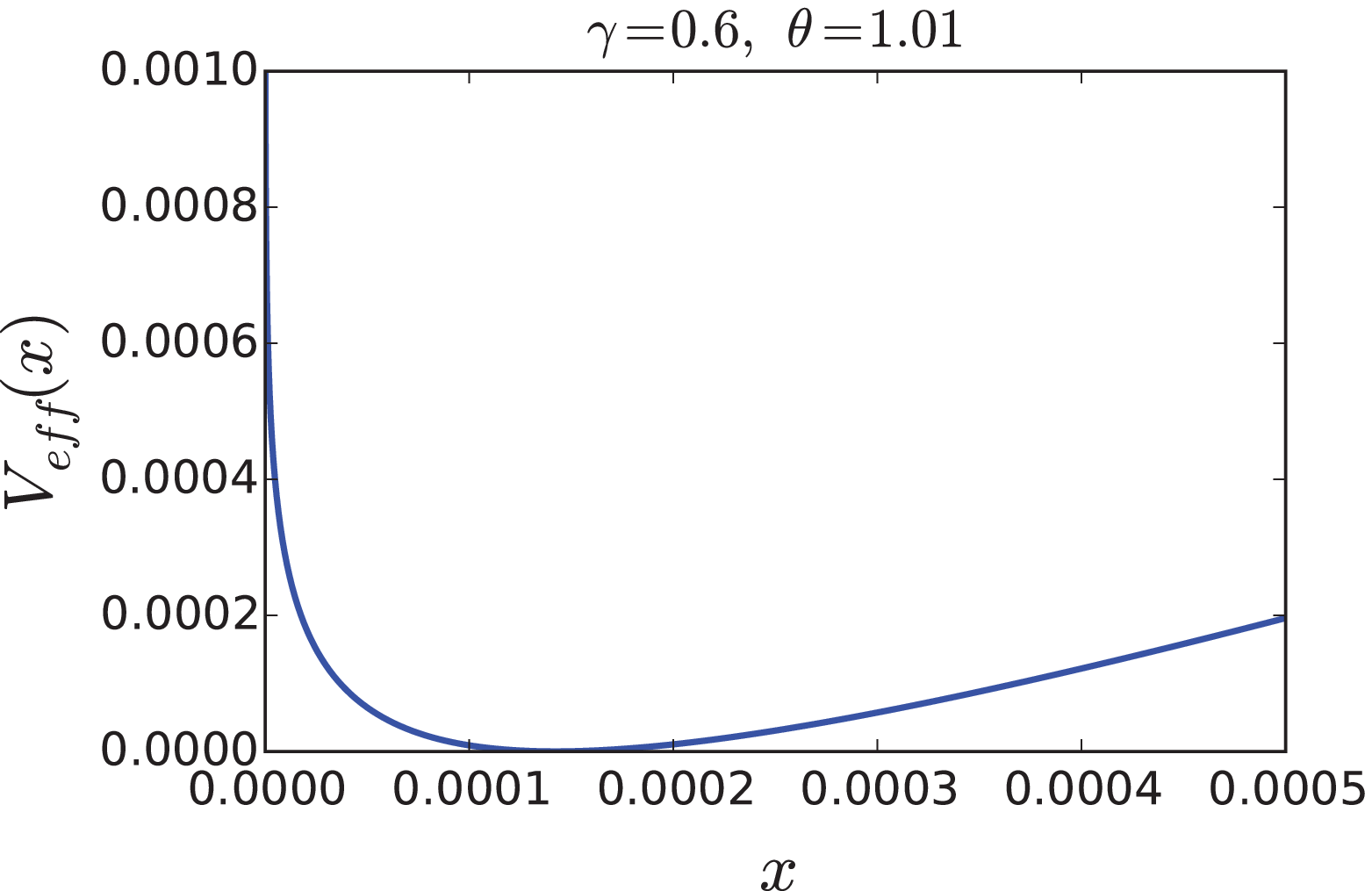}}
    \subfigure{\includegraphics[width=0.325\textwidth]{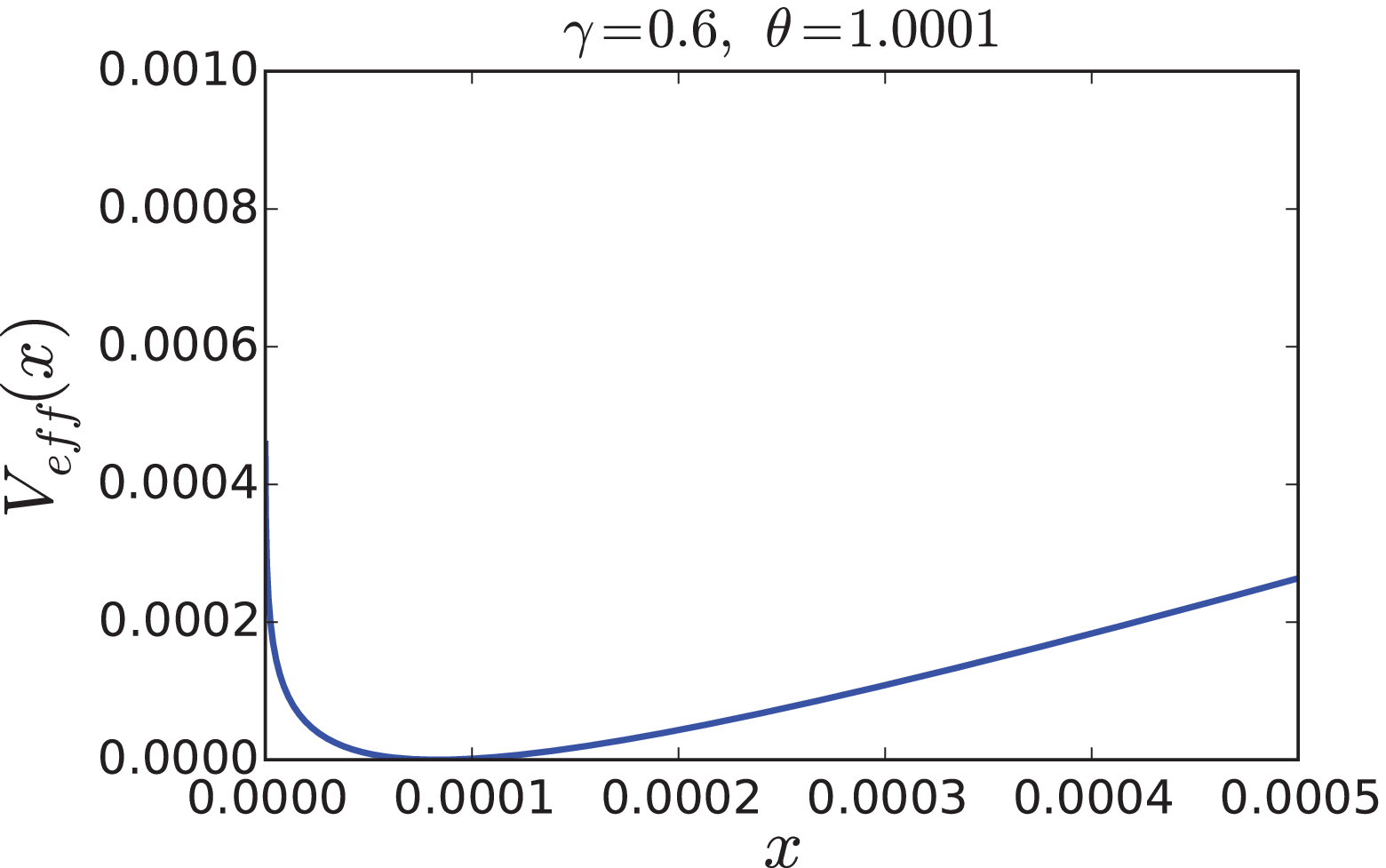}}
    \subfigure{\includegraphics[width=0.325\textwidth]{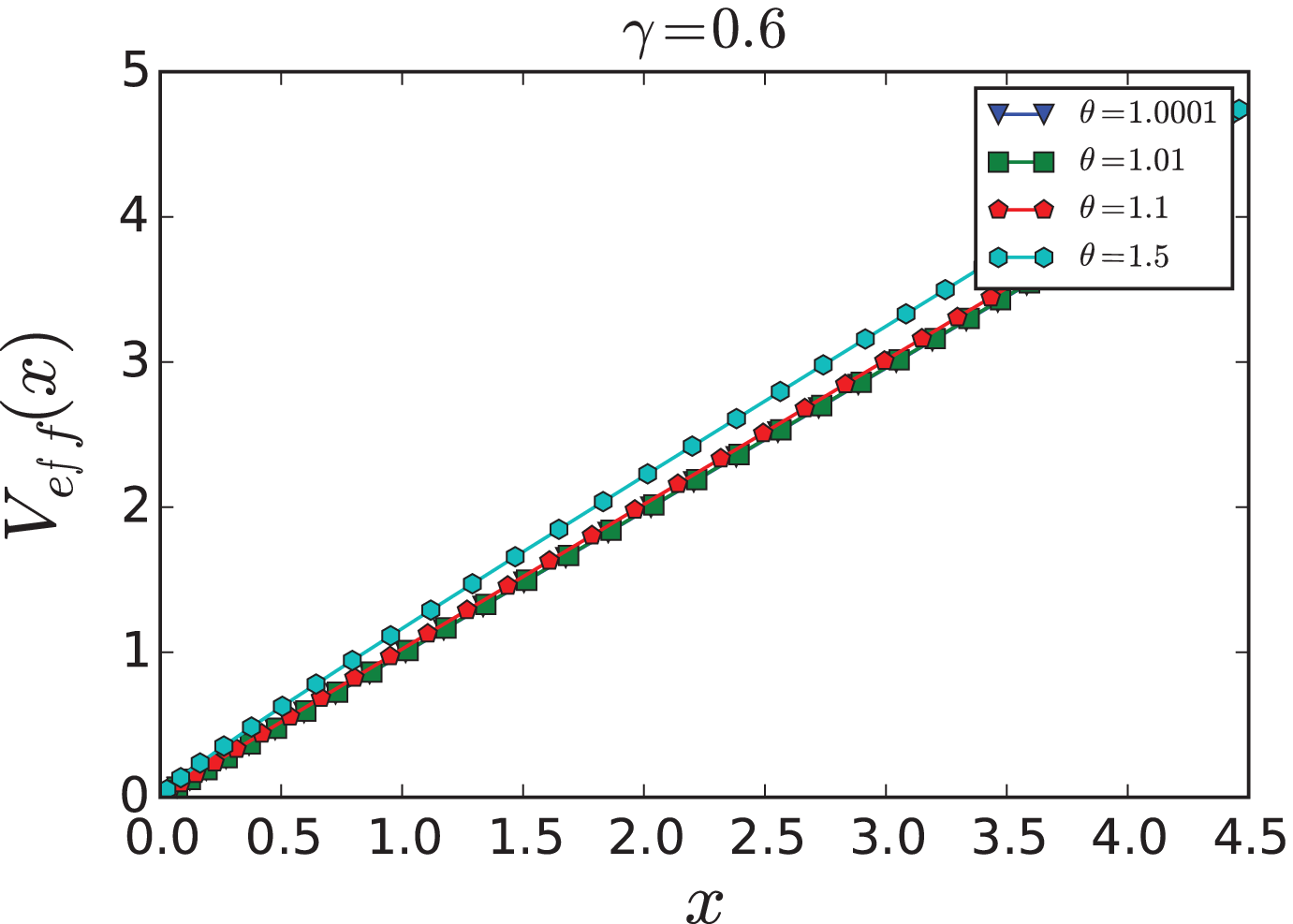}}  
    \caption{(Color online)
Effective potentials close to the origin and over the full support, for $\gamma = 0.6$ and different values of $\theta$. Near the origin, the minima of the non-monotonic effective potential first moves away from the origin and then moves towards the origin as $\theta$ is reduced. Note that the effective potential is monotonic for $\theta =2$ \cite{Alam-Muttalib-Wang-Yadav20}. Also, consistent with the analytical result for $\theta=1$, the non-monotonicity of effective potential near the origin reduces as $\theta\rightarrow 1$.}
    \label{potential_theta_1}
\end{figure*}

In summary, starting with a jpd of the $\gamma$-ensemble with some confining potential $V(x)$, it is possible to map the problem to an MB ensemble ($\gamma=1$), but with an effective potential $V_{eff}(x,\gamma)$ given by Eq. (\ref{V-effective}). Then, the density of the eigenvalues for such an MB ensemble can be obtained using Eq. (\ref{density_hard_edge}). 
We will use this prescription in the following sections to obtain the density of eigenvalues for several different toy models. We will show that one effect of the parameter $\gamma$ is to add non-monotonicity to the effective potential.


\section{Nonmonotonic effective potential for $1<\theta < 2$} \label{sec:3}

As a first step towards understanding the role of the parameter $\gamma$ in the $\gamma$-ensembles, we consider a range of the parameter $\theta$, beyond the value $\theta=2$  considered in detail in \cite{Alam-Muttalib-Wang-Yadav20}. The idea is to show first of all that for certain range of $\theta$, the effective potential can become non-monotonic near the origin. Within that range, the goal is then to choose a particular fixed value of $\theta$ that shows a significant non-monotonicity and systematically study the effective potential as well as the eigenvalue density  as a function of $\gamma$. This would allow us to focus on the role of $\gamma$ in the $\gamma$-ensembles. We will restrict ourselves to the case $\gamma <1$, which is expected to be relevant for disordered quantum conductors.

Figure \ref{potential_theta_1} shows the effective potentials near the origin for $\gamma=0.6$ and a range of values for $\theta$ between $1$ and $2$. We have shown in \cite{Alam-Muttalib-Wang-Yadav20} that the effective potential for $\theta=2$ monotonically goes to zero at the origin. As $\theta$ is reduced from $2$, the effective potential develops a non-monotonicity. The minima of the effective potential gradually becomes deeper and moves away from the origin. Later as $\theta$ moves closer to $1$, the depth of the minima of the effective potential decreases and the minima shifts closer to the origin. Thus with decreasing non-monotonicity, we expect the effective potential to become linear for $\theta=1$ as predicted by Eq. (\ref{V_eff_theta_1}). We have also verified this expected analytical results for $\gamma >1$ case.

\begin{figure}
\begin{center}
\includegraphics[width=0.4\textwidth]{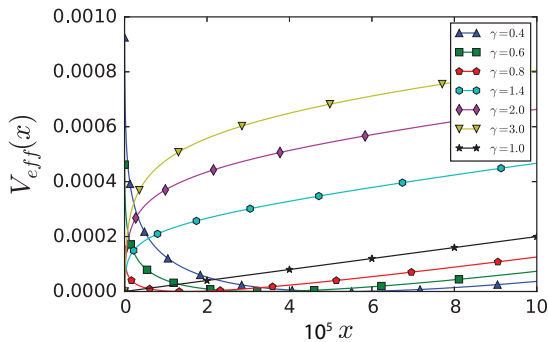}
\end{center}
\caption{
(Color online) 
Effective potential near the origin for different $\gamma$, $V(x)=2x$ and $\theta = 1.0001$.}
\label{potential_near_origin}
\end{figure}

Figure \ref{potential_theta_1} suggests that even for  $\theta$ close enough to $\theta=1$,  the effect of $\gamma$ on the non-monotonicity could be observable. We therefore choose $\theta = 1.0001$ and a linear external potential, $V(x)=2x$. Figure \ref{potential_near_origin} shows the effective potential for different values of $\gamma$, where we include $\gamma > 1$ as well to show that the results are qualitatively different.

Note that the limit $\theta=1$ is identical to the well-known $\beta$-ensembles with $\beta=\gamma+1$. Analytical results for such Laguerre $\beta$-ensembles obtained in Appendix A suggest that the non-monotonicity of the effective potential should disappear  at $\theta = 1$. The present formalism allows us to consider the $\theta \to 1$ limit and thereby obtain the effective potential as well as the density for $\beta$-ensembles for arbitrary $\beta$, as shown in the Appendix.

\begin{figure}
\begin{center}
\includegraphics[width=0.4\textwidth]{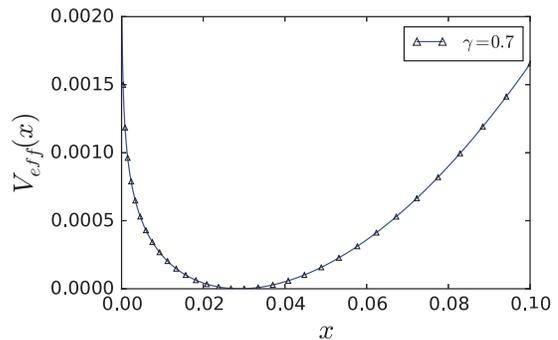}
\end{center}
\caption{
(Color online) 
Effective potential near the origin for quadratic potential $V(x)=0.2x^2$, $\gamma=0.7$ and $\theta = 1.0001$.}
\label{potential_quadratic}
\end{figure}

\begin{figure}[h]
\begin{center}
\includegraphics[width=0.48\textwidth]{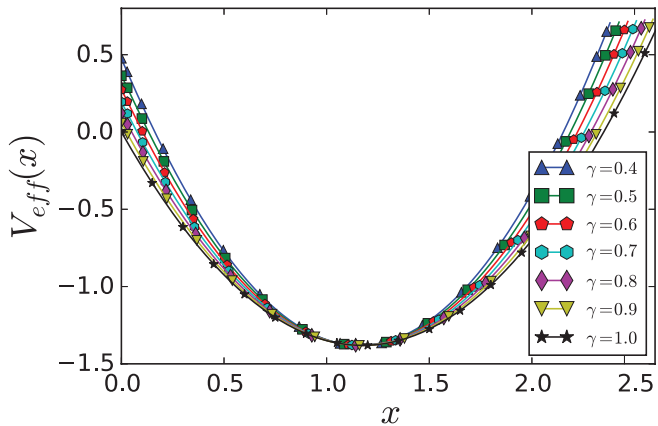}
\end{center}
\caption{
(Color online) 
Effective potential for $\theta = 1.2$, $V(x)=x^2-2.35x$ and different $\gamma$.}
\label{potential_transition}
\end{figure}

\begin{figure*}
    \centering
    \subfigure{\includegraphics[width=0.325\textwidth]{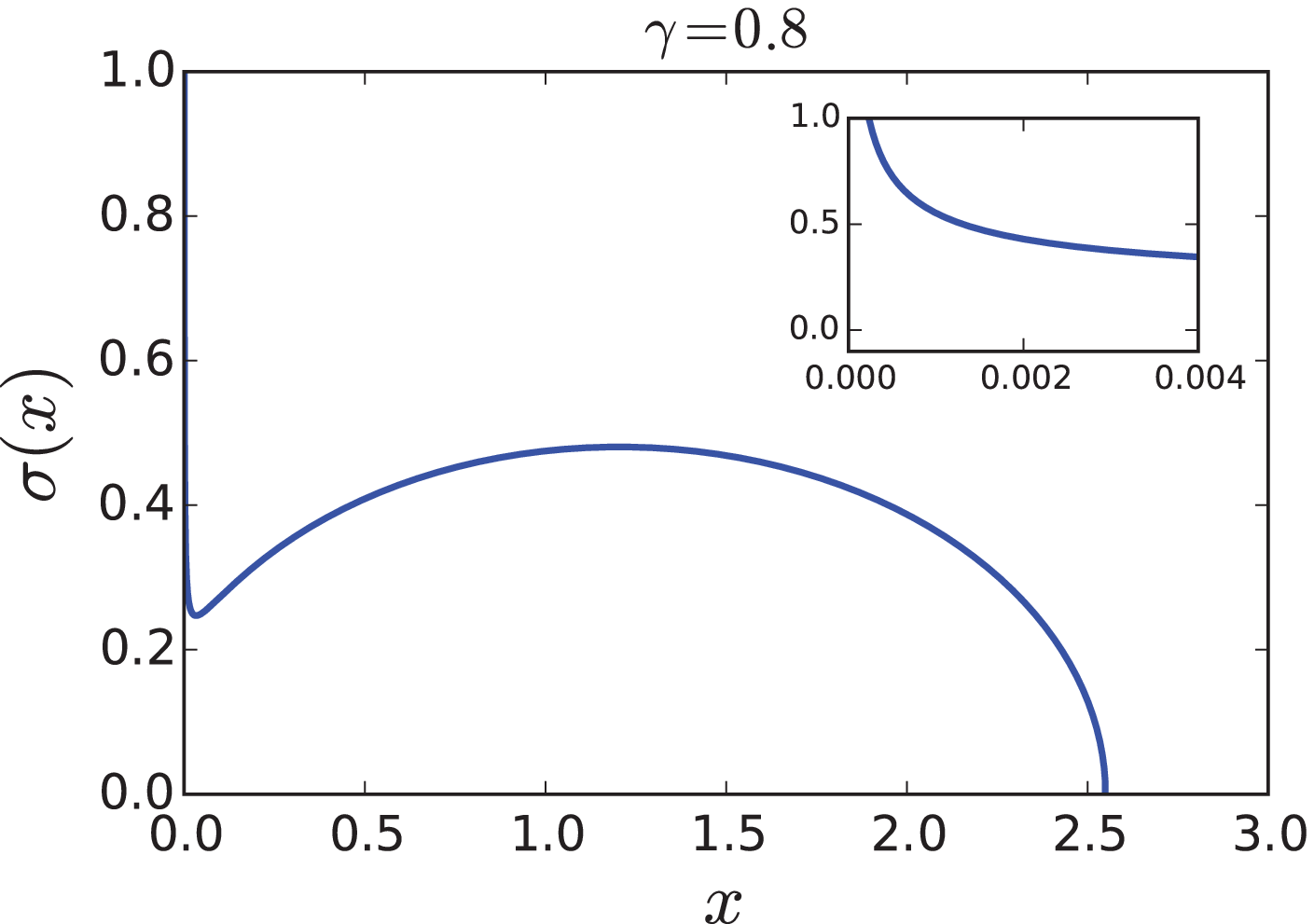}}
    \subfigure{\includegraphics[width=0.325\textwidth]{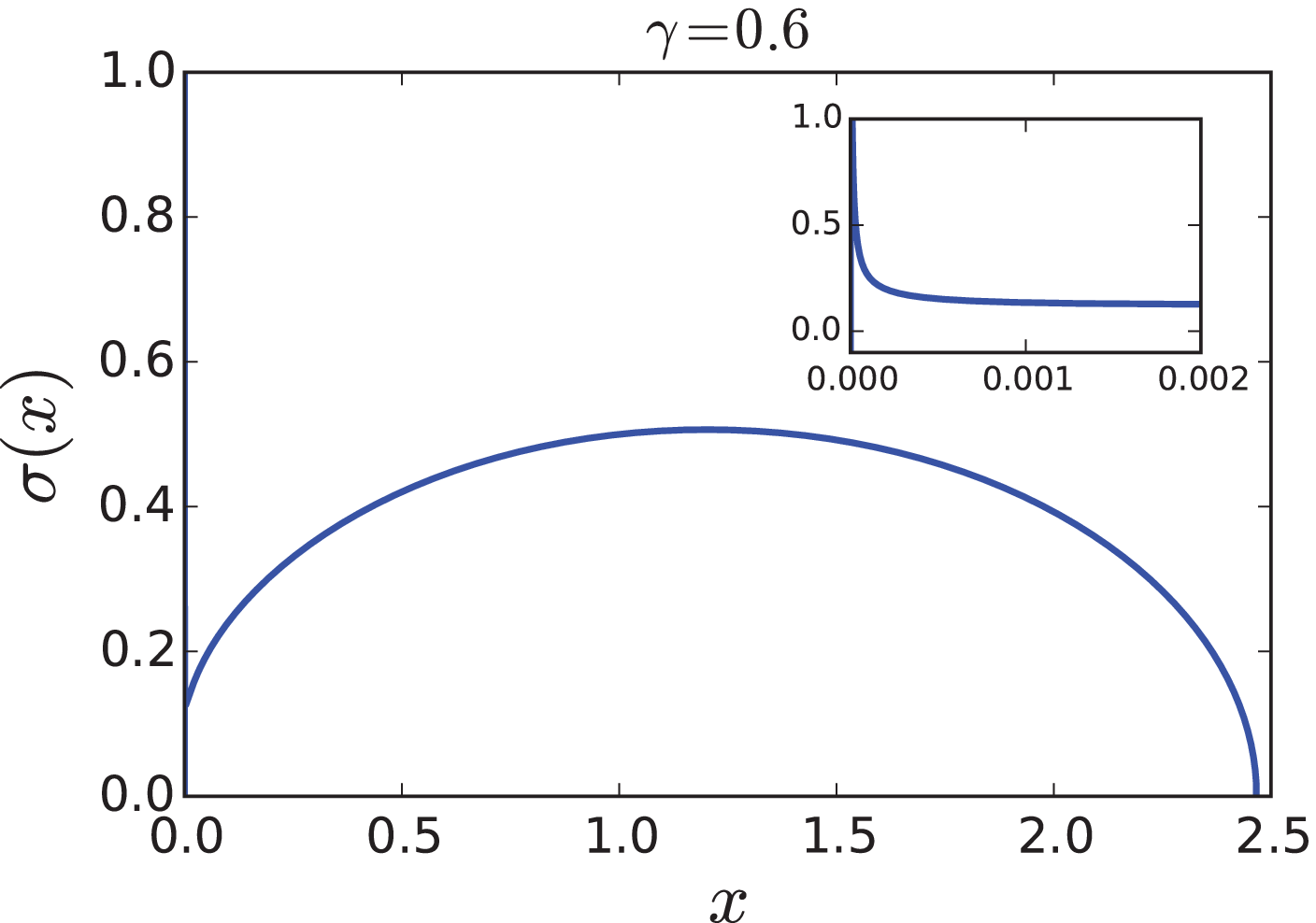}}  
    \subfigure{\includegraphics[width=0.325\textwidth]{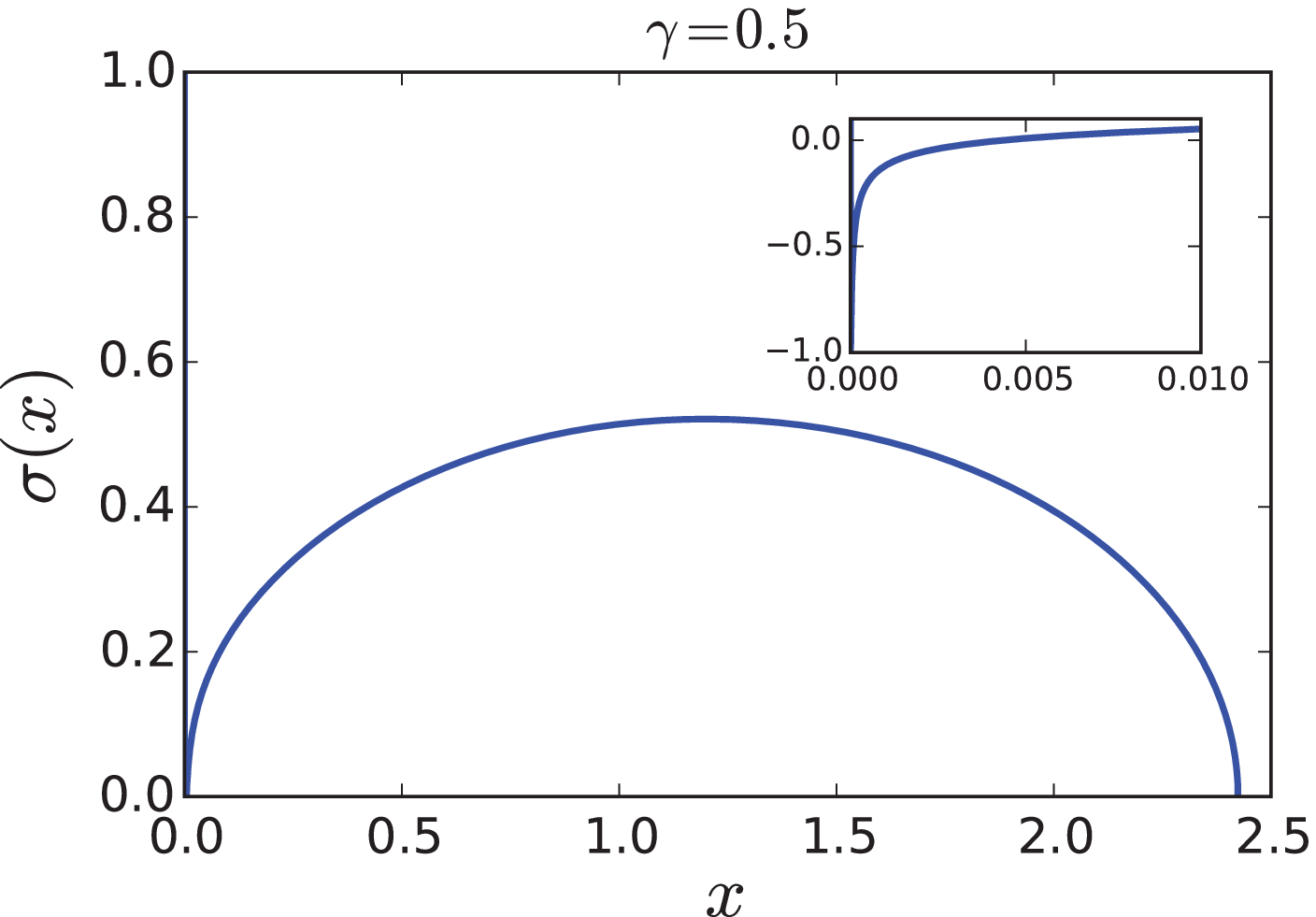}}
    \subfigure{\includegraphics[width=0.325\textwidth]{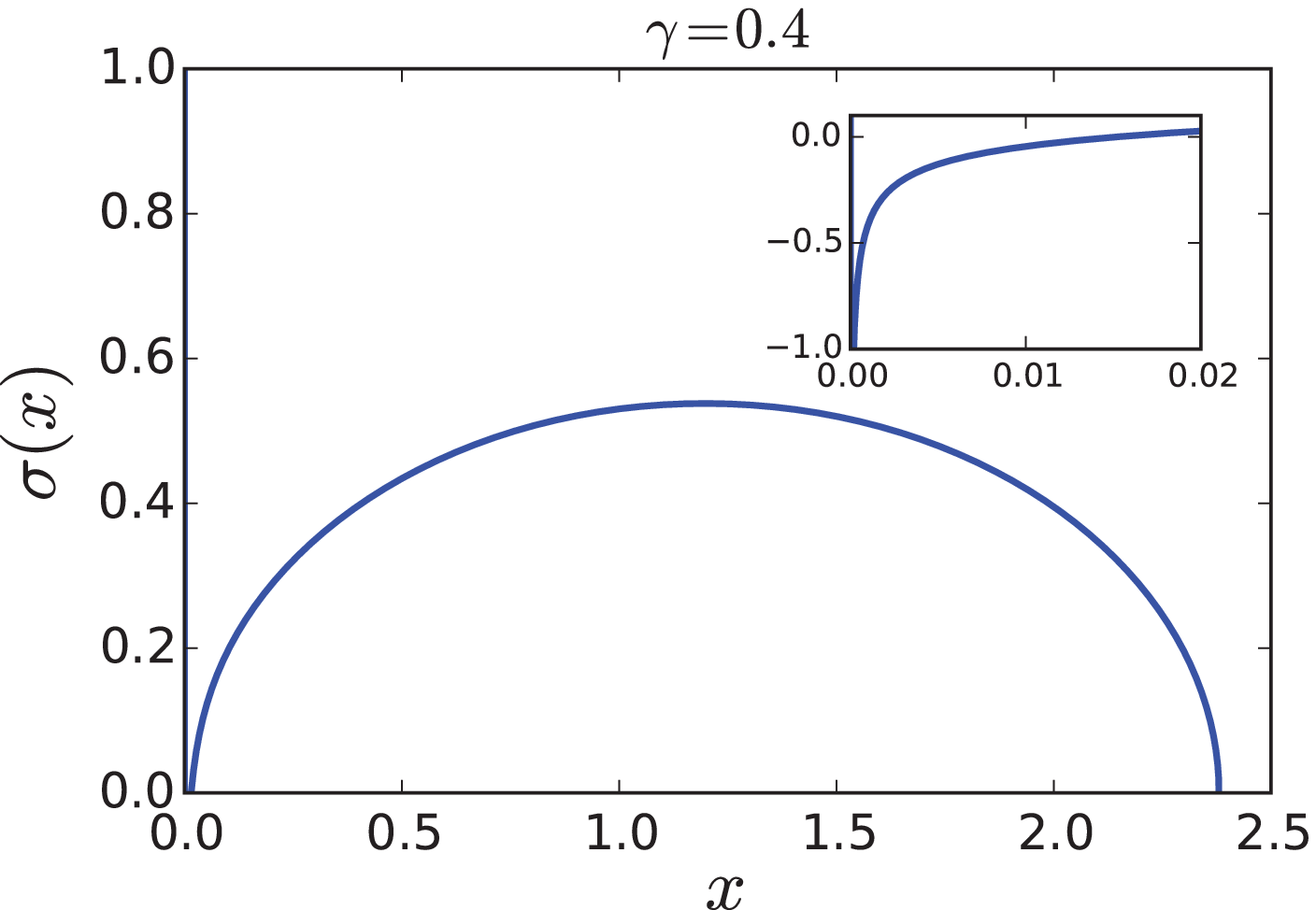}}        
    \subfigure{\includegraphics[width=0.325\textwidth]{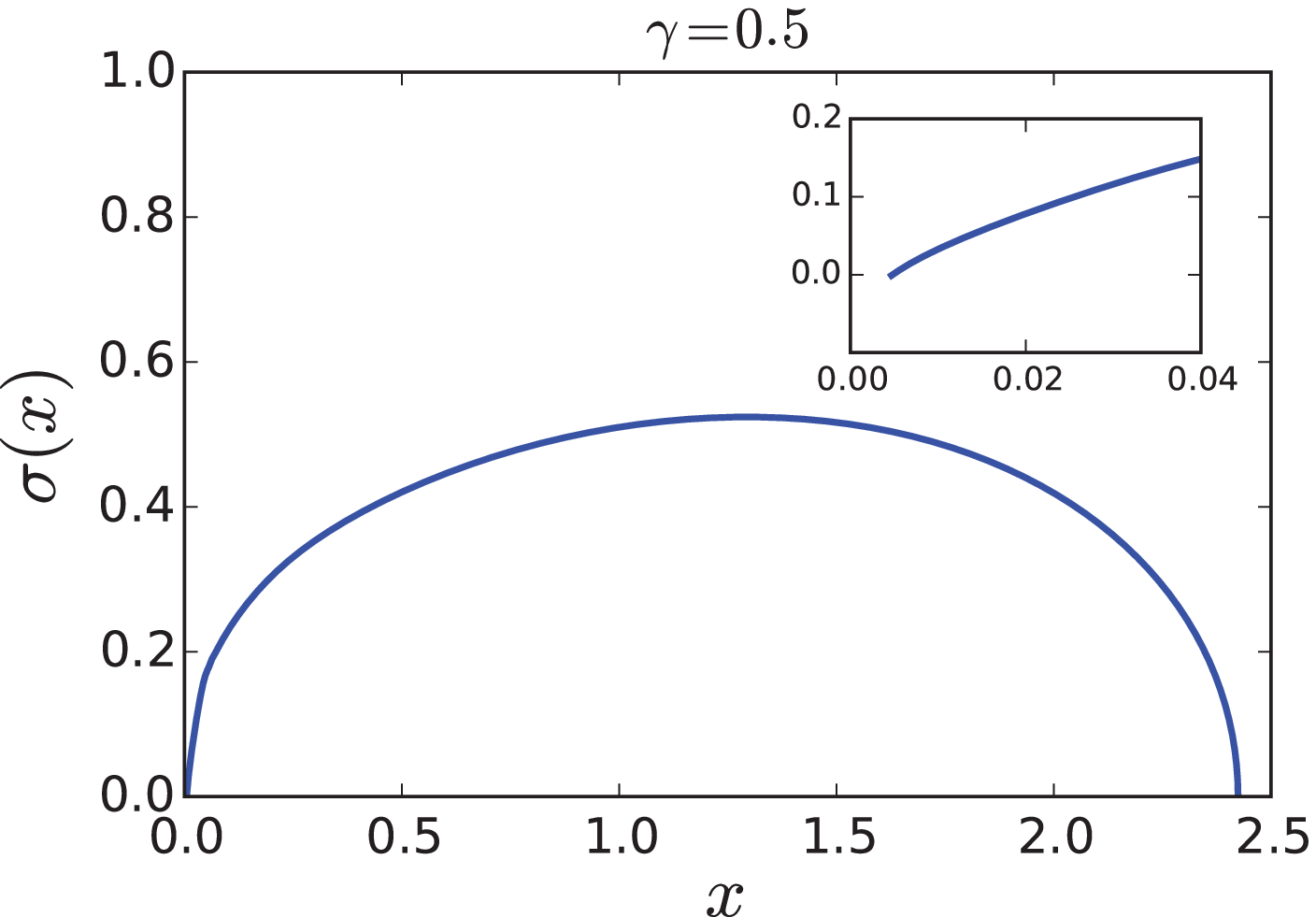}}
    \subfigure{\includegraphics[width=0.325\textwidth]{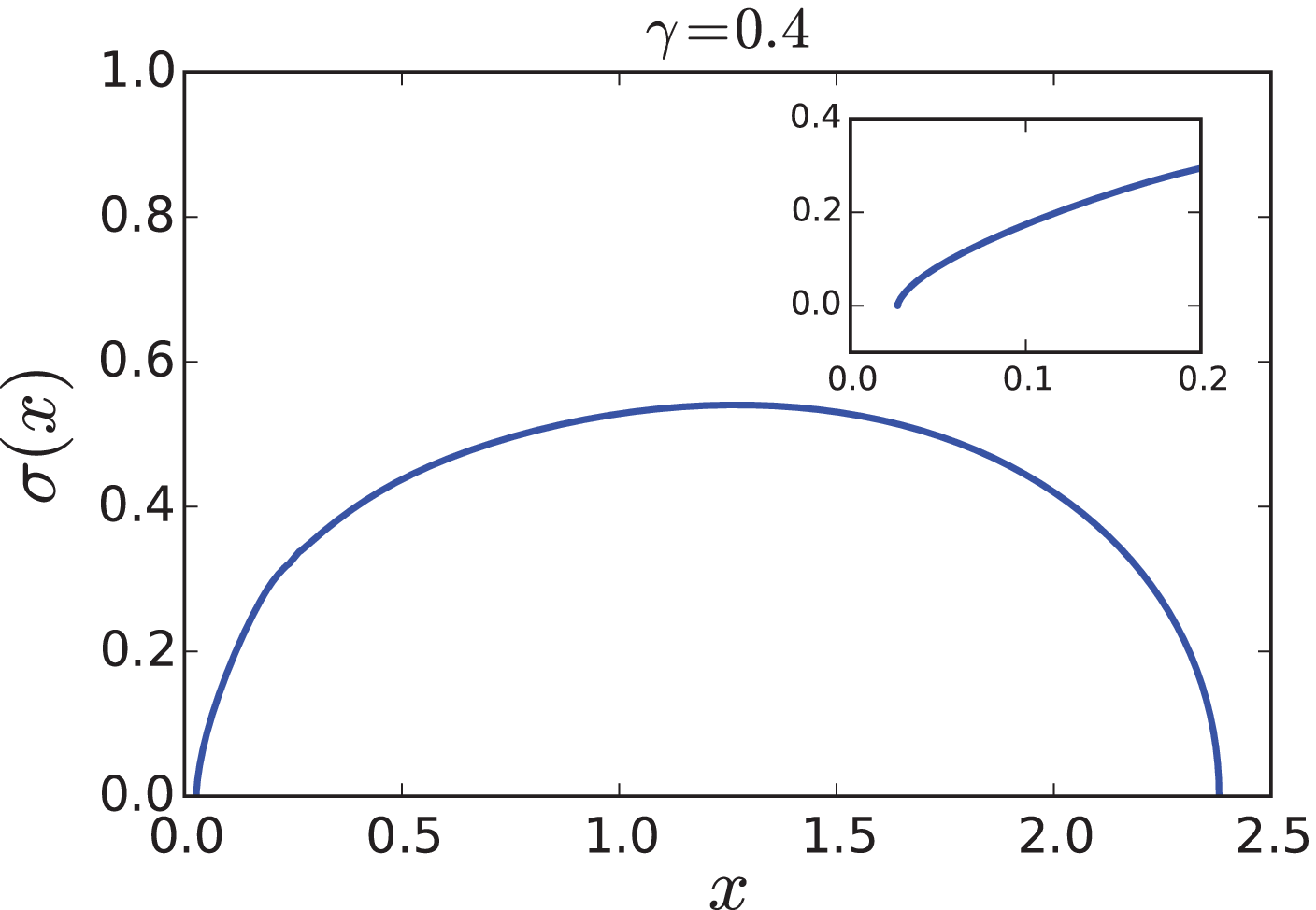}}    
    \caption[]{(Color online)
The eigenvalue density, for $\theta = 1.2$, $V(x)=x^2-2.35x$, and different values of $\gamma$. Inset shows the corresponding density near the origin. For $\gamma = 0.5$ and $0.4$, the hard-edge eigenvalue densities become negative near the origin,  implying that the assumption of hard-edge support is wrong and true density has a soft-edge support. The last two panels (the small kinks in the density are numerical artifacts and go away with finer grid)  show the true eigenvalue density for $\gamma = 0.5$ and $0.4$ with the soft-edge support. }
    \label{density_transition}
\end{figure*}


\begin{figure*}
    \centering
      \subfigure{\includegraphics[width=0.325\textwidth]{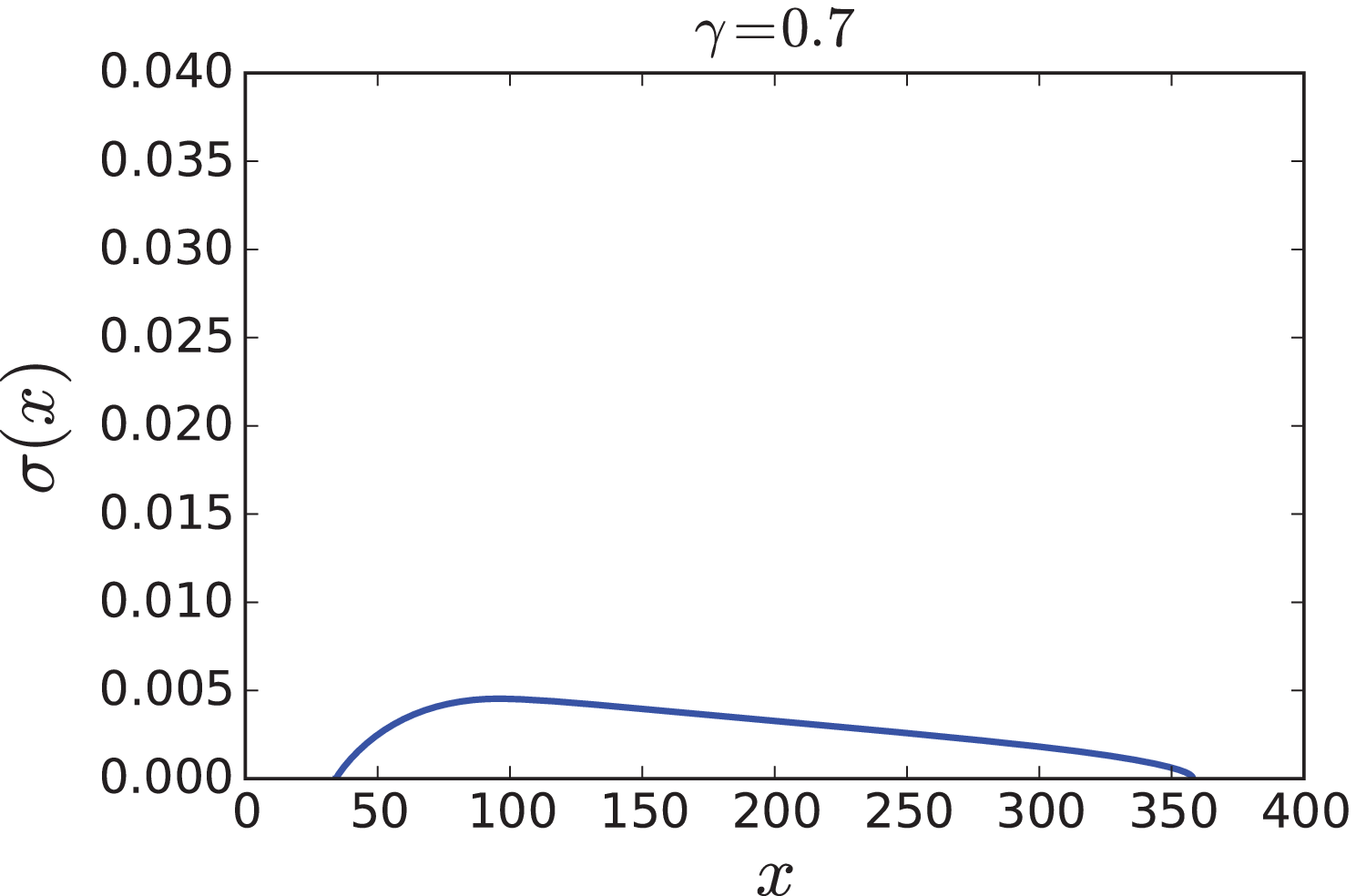}}  
       \subfigure{\includegraphics[width=0.325\textwidth]{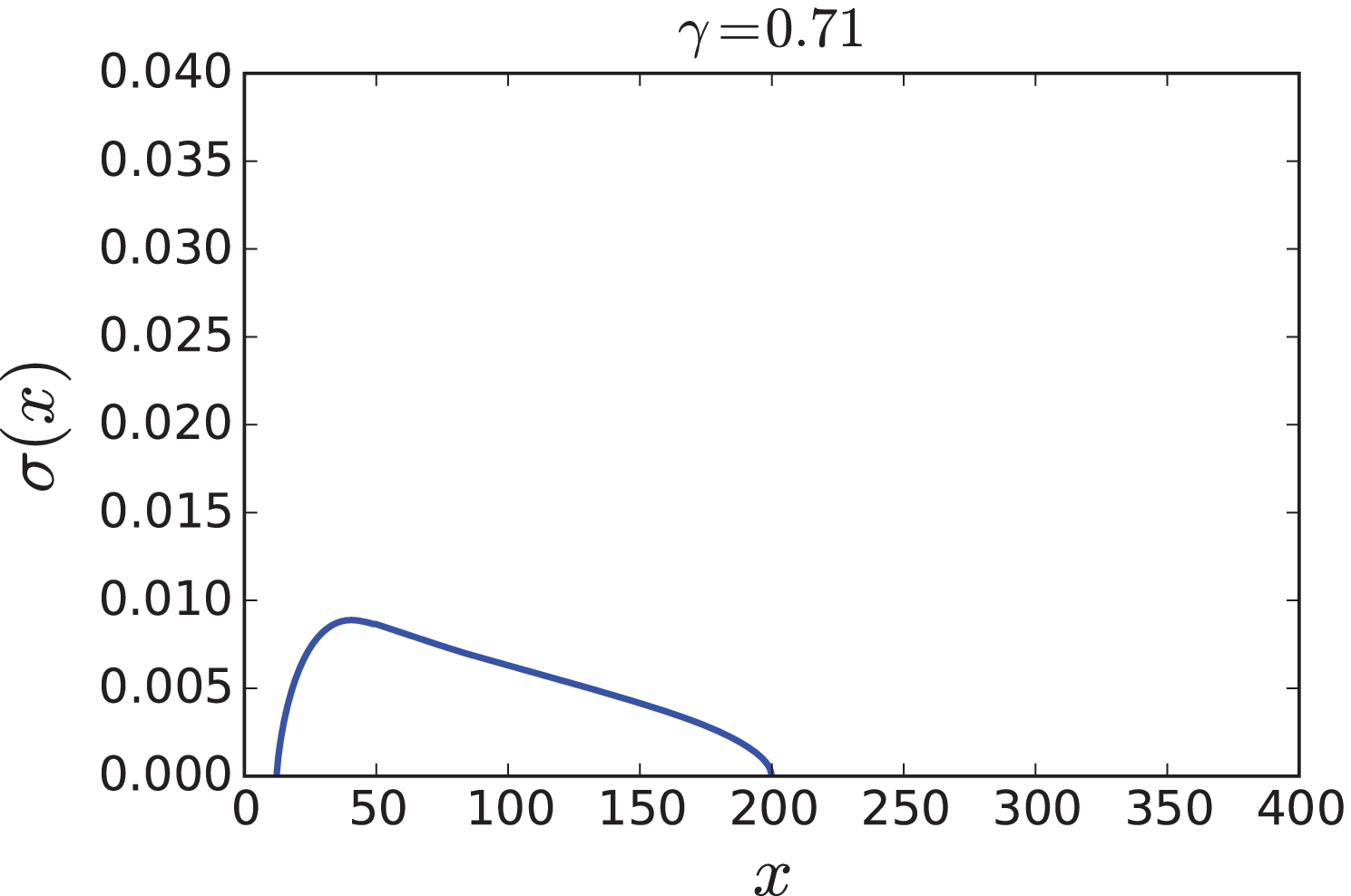}}
        \subfigure{\includegraphics[width=0.325\textwidth]{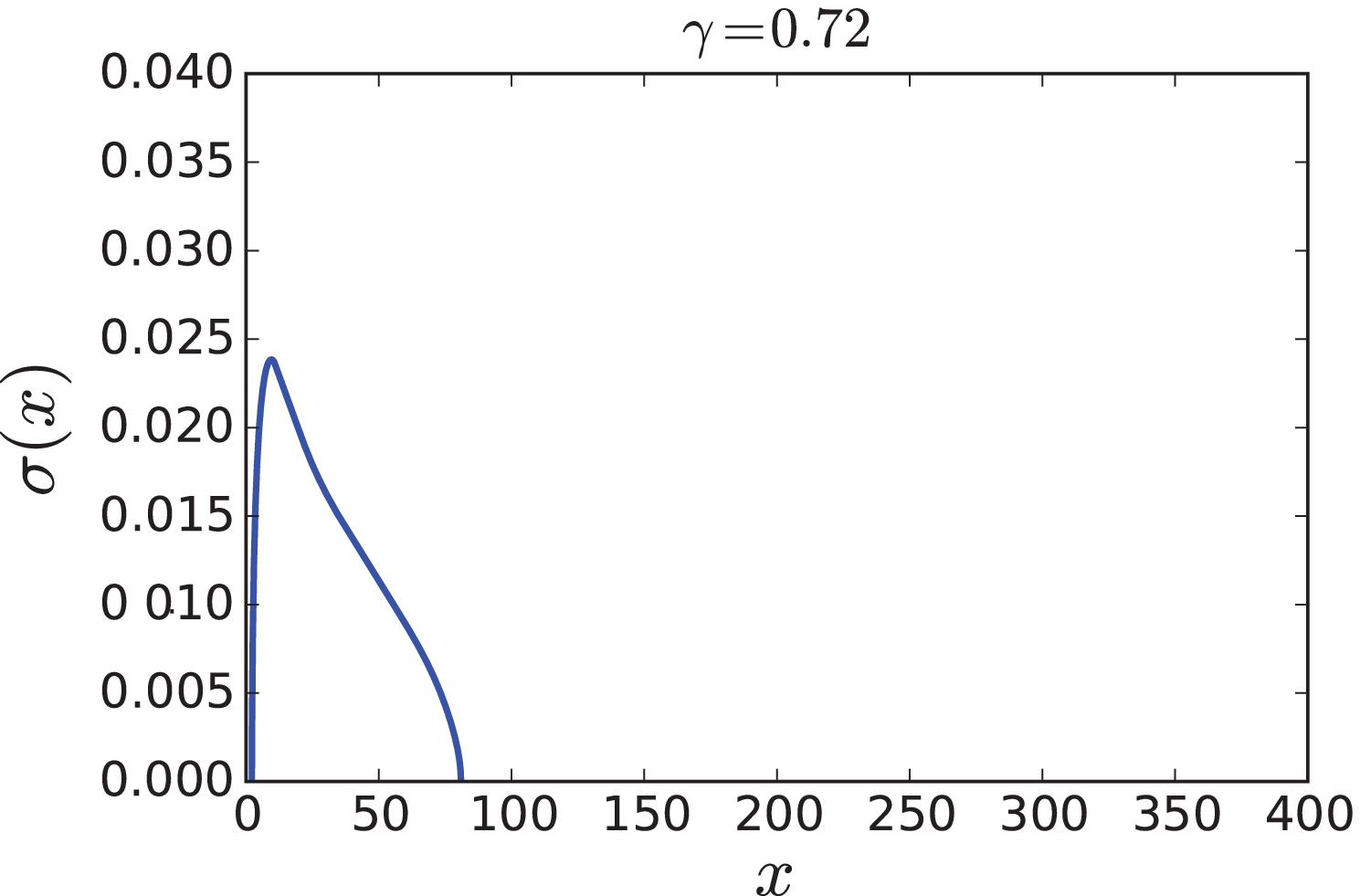}}
         \subfigure{\includegraphics[width=0.325\textwidth]{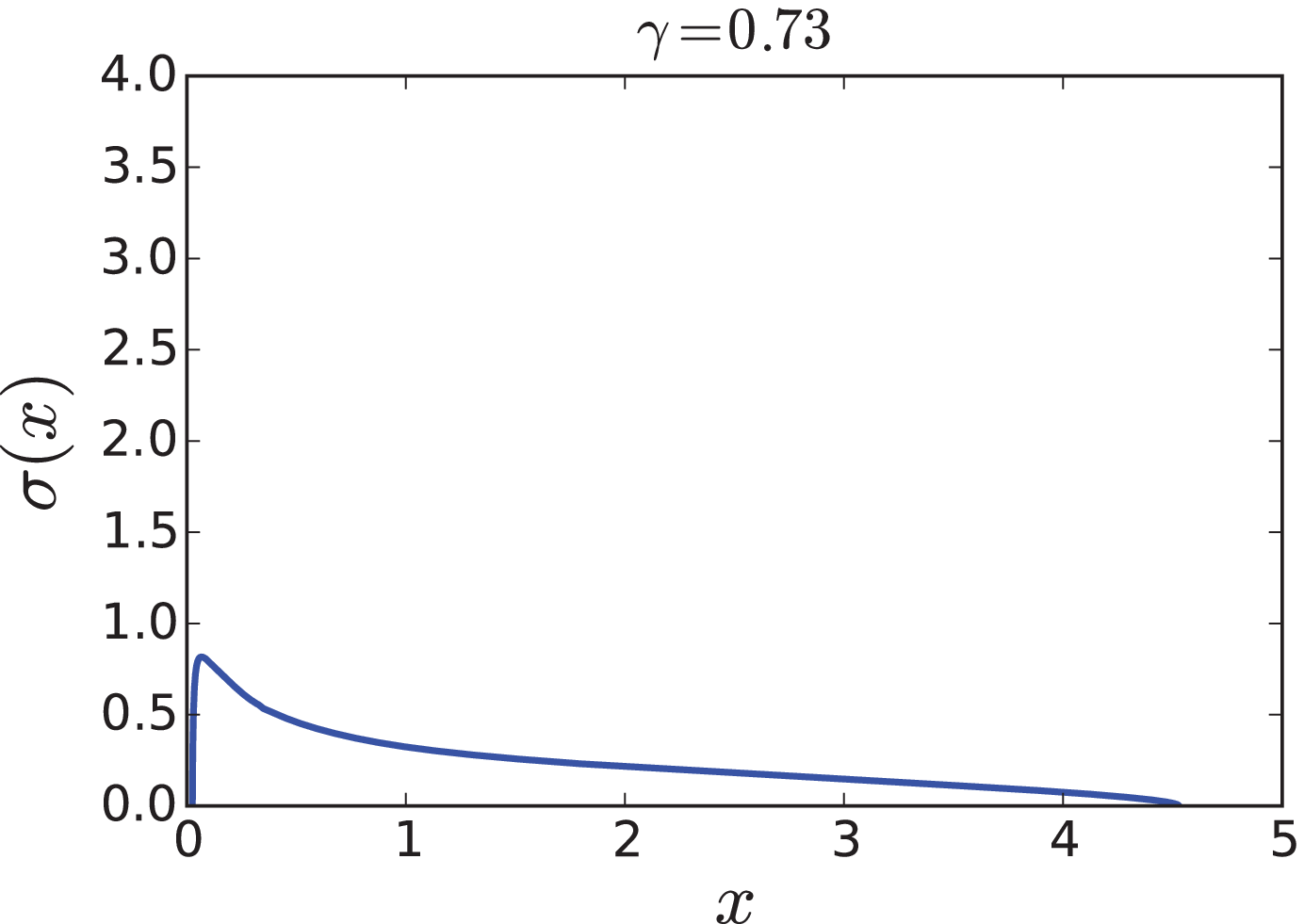}}
     \subfigure{\includegraphics[width=0.325\textwidth]{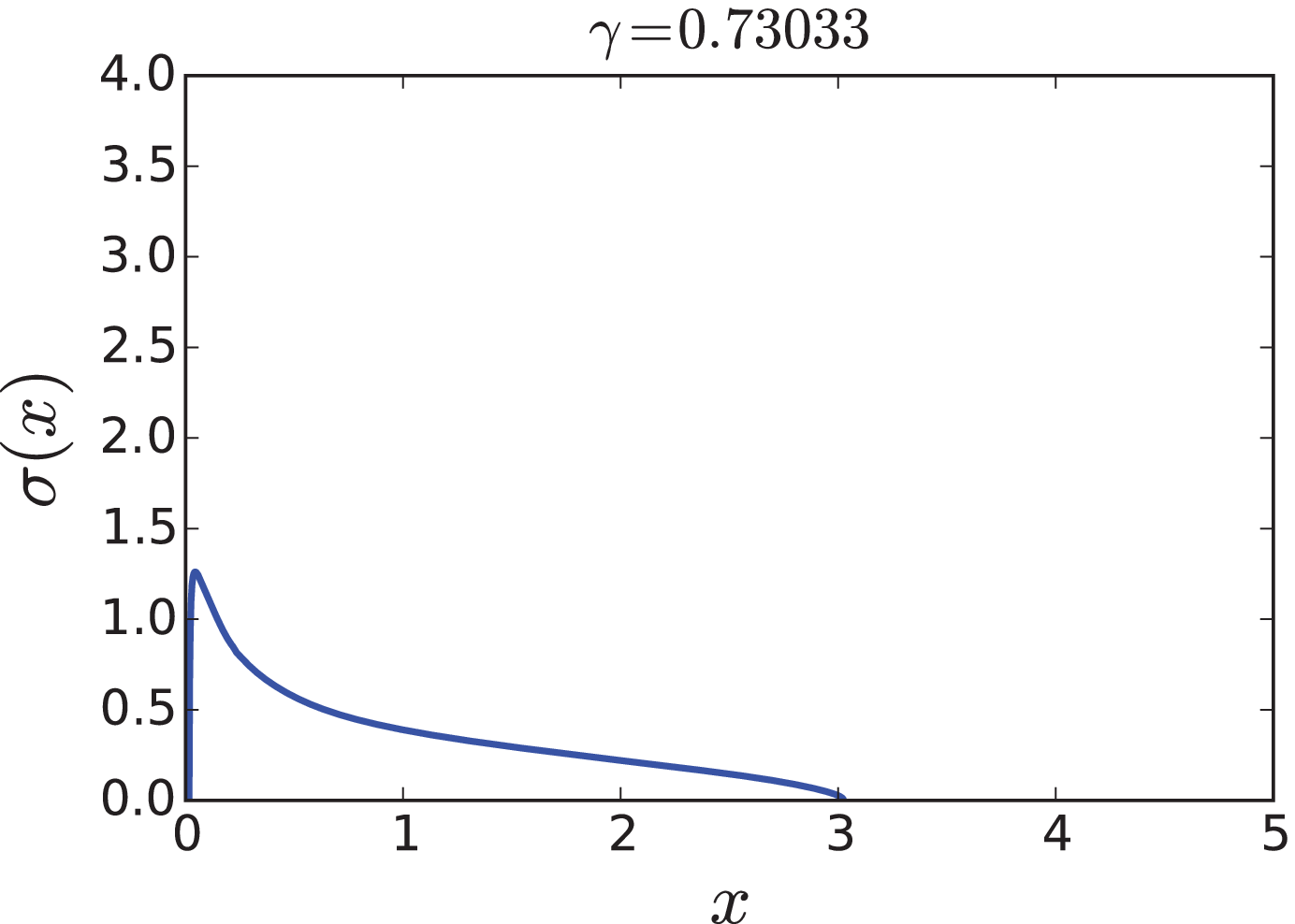}}
      \subfigure{\includegraphics[width=0.325\textwidth]{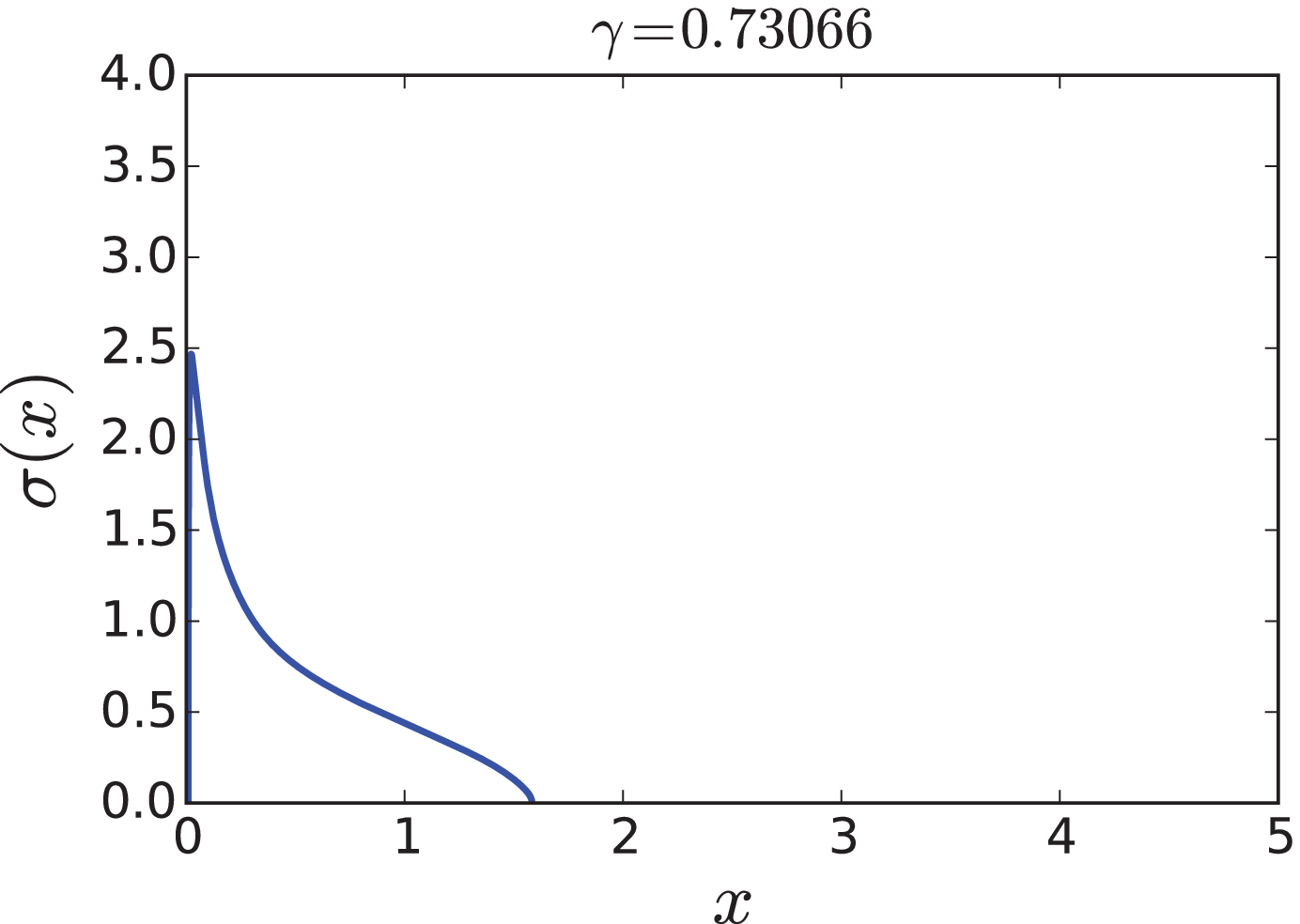}}
       \subfigure{\includegraphics[width=0.325\textwidth]{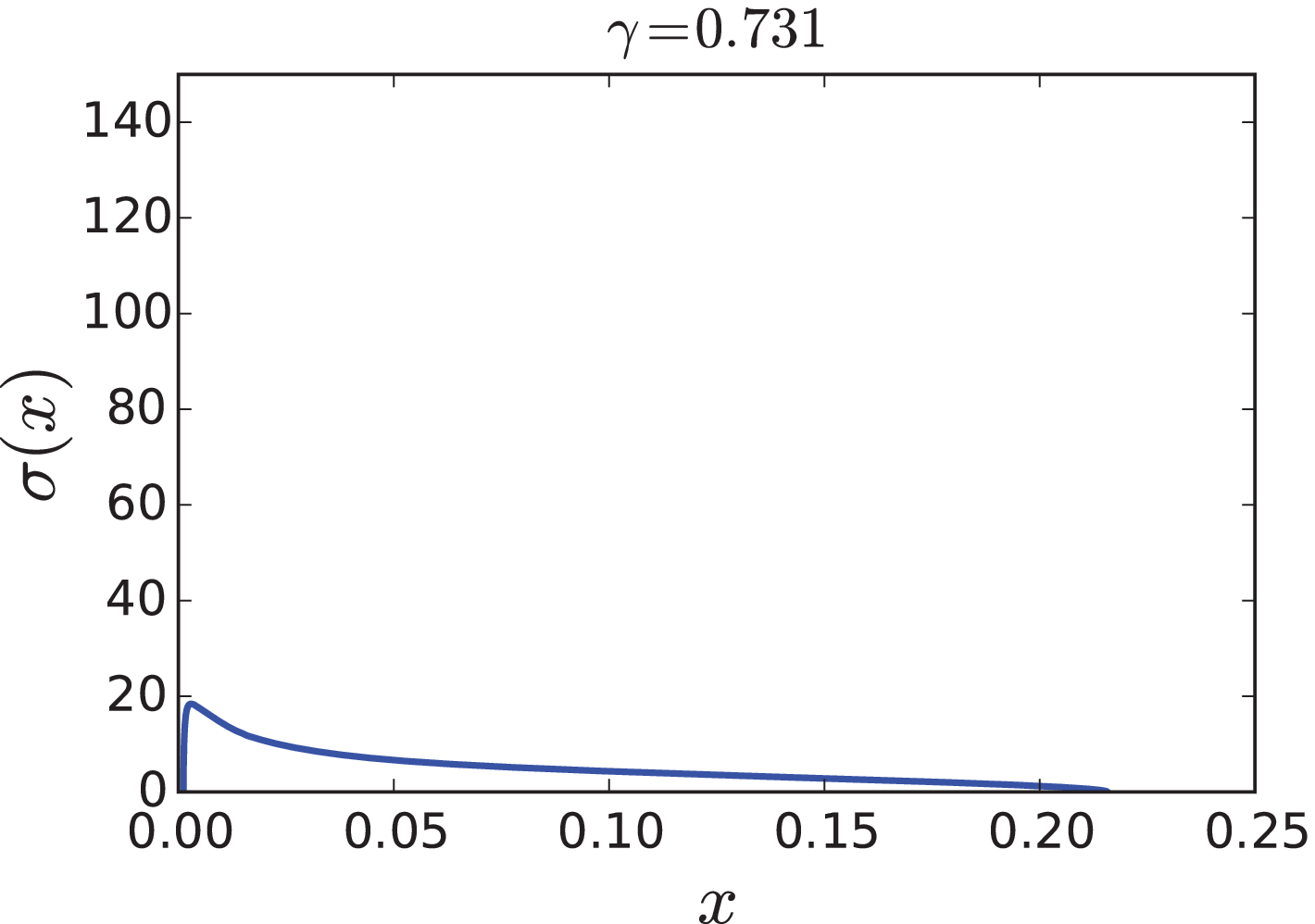}}
        \subfigure{\includegraphics[width=0.325\textwidth]{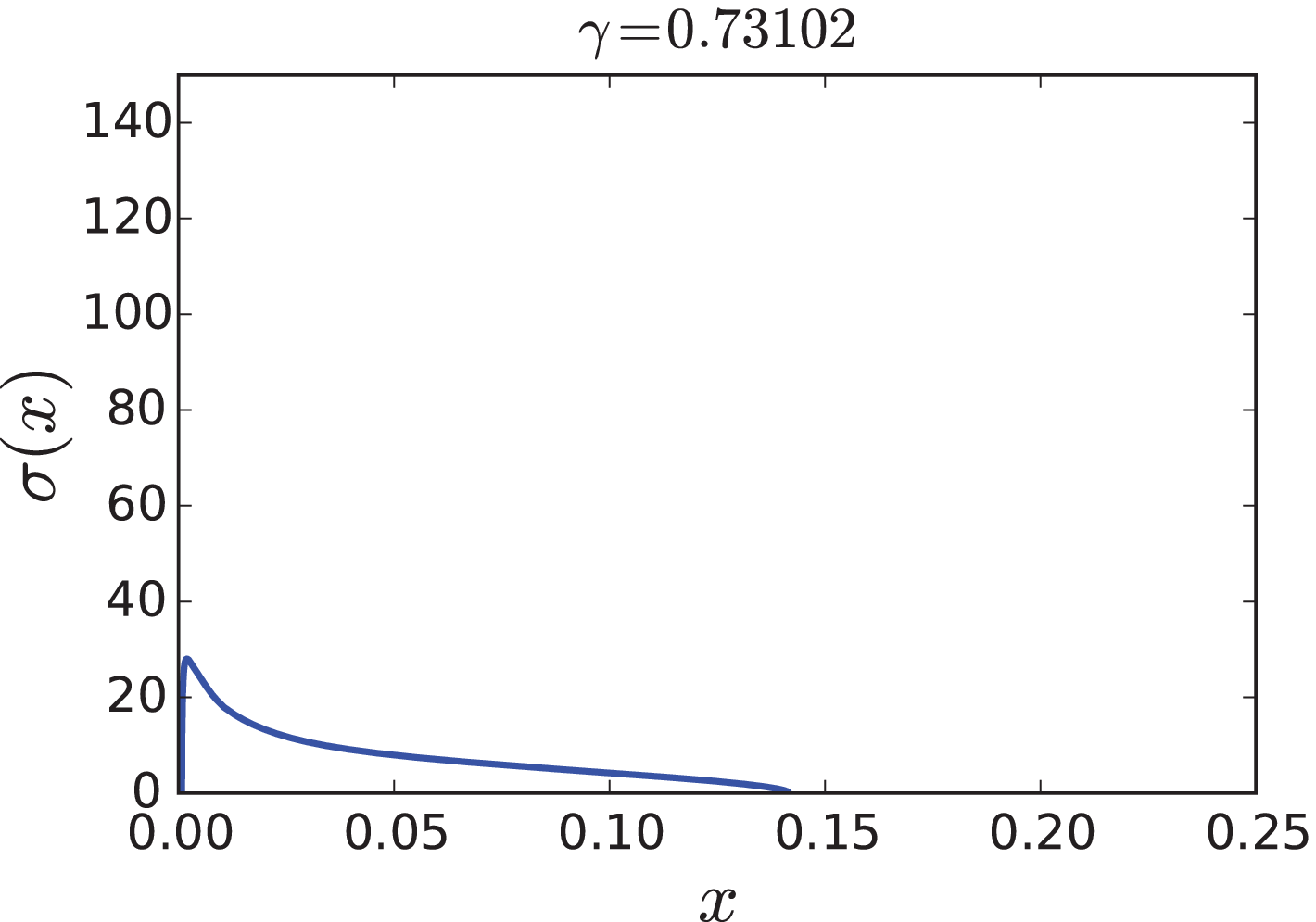}} 
     \subfigure{\includegraphics[width=0.325\textwidth]{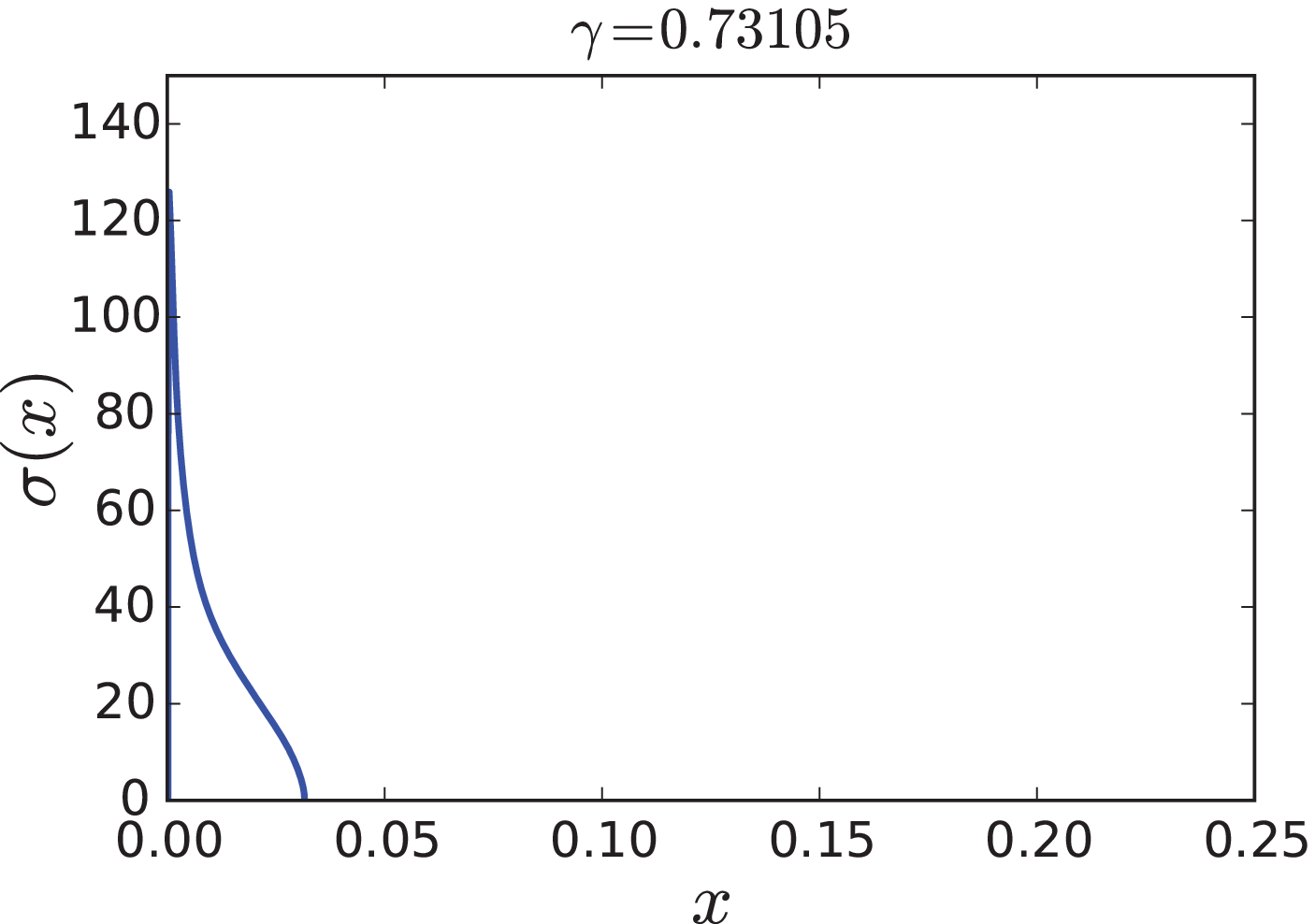}}        
 \caption{(Color online)
The eigenvalue density, for $\theta = 1.8$, $V(x)=\frac{a\gamma}{1+\ln\frac{1-\gamma}{\gamma}}x-\frac{1}{2}\ln(\sinh 2\sqrt{x})$ with $a=0.01$, and different values of $\gamma$. All densities have soft-edge support.}
    \label{density_MMW_sigmoid}
\end{figure*}

To explore how the non-monotonicity changes with the single-particle potential, we consider the $\gamma$-ensemble with a quadratic single-particle potential $V(x)=\alpha x^2$, $\gamma = 0.7$ and $\theta \rightarrow 1$. We choose $\alpha = 0.2$ so that the potential is much weaker near the origin compared to the linear potential. Figure \ref{potential_quadratic} shows that the minima of the effective potential is shifted significantly away from the origin and is deeper compared to the effective potentials in Fig. \ref{potential_near_origin}.


\section{Hard-edge to soft-edge transition for eigenvalue density} \label{sec:4}

In the previous section  we showed that the effect of decreasing the exponent $\gamma$ from $1$  in the $\gamma$-ensembles with either a linear or a quadratic single-particle potential is equivalent to adding a non-monotonicity in the effective potential for the corresponding MB ensembles.  It has been shown in \cite{Claeys-Romano14} that such a minima in confining potential, if deep enough, can produce a transition from a diverging eigenvalue density at the hard-edge to a non-diverging density.  However, the non-monotonic effective potentials we have computed in these cases for different $\gamma$ and different $\theta$ are not sufficient to produce the hard-edge to soft-edge transition in the eigenvalue density. In this section we show that starting with a given non-monotonic potential of the form $V(x)= x^2-\rho x$, with fixed $\rho=2.35$ for which the density is still diverging near the origin, changing $\gamma$ alone is sufficient to produce such a transition. Note that this is qualitatively different from the CR model \cite{Claeys-Romano14}, where a transition is obtained by changing the non-monotonicity parameter $\rho$ in the single-particle potential, while we keep $\rho$ fixed, and change the two-particle interaction parameter $\gamma$ which is expected to be related to the strength of disorder in a three-dimensional quantum conductor.

 We choose the interaction parameter $\theta = 1.2$ because the results from Fig. \ref{potential_theta_1} suggest that for a given $\gamma$, the non-monotonicity in the effective potential is qualitatively the largest for $\theta$ between $1.1$ and $1.5$.  
 For all $\gamma < 1$, we begin with the assumption that the support of density is hard-edge (i.e. the support starts at the origin)) and we use the hard-edge formalism to compute the eigenvalue density. If for some $\gamma < 1$, our assumption of hard-edge support for density is wrong  and the actual support is soft-edge (i.e. the support starts away from the origin) then the hard-edge formalism gives a negative (unphysical) density near origin. In that case, we switch to the soft-edge formalism described in Appendix B and compute the non-negative density with soft-edge support. As the $\gamma$ decreases from $1$, the effective potential increases (becomes more and more non-monotonic) near origin, as shown in Fig. \ref{potential_transition}. For some critical value of $\gamma$ between $0.5$ and $0.6$, this added non-monotonicity in the effective potential brings about the hard-edge to soft-edge transition in the density, see Fig. \ref{density_transition}. As $\gamma$ is reduced further, the soft-edge of the support of the density near origin moves further and further away from origin, increasing the gap in the spectrum.




\section{Phenomenological model for 3D disordered conductors} \label{sec:5}

In this section we consider a phenomenological model based on results from \cite{Markos-Muttalib-Woelfle-Klauder04, Markos-Muttalib-Wolfle05}. We will restrict ourselves to 3D only;  for a brief discussion of how the dimensionality enters the current formulation, see Appendix C. The jpd for the ensemble is given by
 \cite{Markos-Muttalib-Woelfle-Klauder04, Markos-Muttalib-Wolfle05, Klauder-Muttalib99, Gopar-Muttalib02, Douglas-Markos-Muttalib14}
\begin{equation}
\begin{aligned}
  &p(\{x_i\};\gamma) \propto \prod_{i=1}^Nw(x_i,\gamma)\prod_{i<j}|x_i-x_j||s(x_i)-s(x_j)|^{\gamma},\\
  &w(x,\gamma) = e^{-N V(x,\gamma)}.
\end{aligned}
\label{DMPK_3D_MMW}
\end{equation}
where $s(x)=\sinh^2 \sqrt{x}$.

The Joukowsky transformation for the interaction term,  $|\sinh^2\sqrt{x_i}-\sinh^2\sqrt{x_j}|$, is not available and hence the explicit numerical solution to the RH problem associated with this jpd can not be obtained. Fortunately, the $x^{\theta}$ interaction term in $\gamma$ ensembles with $\theta = 1.8$ and the $\sinh^2\sqrt{x}$ interaction term in Eq. (\ref{DMPK_3D_MMW}) have very similar qualitative behavior over a reasonable range of support for the eigenvalue density. Thus, we can use the $\gamma$-ensemble interaction term with $\theta = 1.8$ as a solvable toy model. The single-particle potential 
$V(x,\gamma)$  has a dominant linear dependence on $x$ in the strongly disordered regime, whose strength depends on the parameter $\gamma$. It also includes a logarithmic part arising from a Jacobian of transformation. In the strong disorder regime, the total single-particle potential is given by \cite{Markos-Muttalib-Wolfle05}  
\be
V(x,\gamma)=\Gamma x-\frac{1}{2}\ln(\sinh 2\sqrt{x}),
\label{Vofgamma}
\ee
where the coefficient $\Gamma$ depends on disorder, but its functional relationship with the two-particle interaction parameter $\gamma$ is not known in general. The relationship has been discussed only in the strongly disordered insulating regime \cite{Markos-Muttalib-Wolfle05} where $\Gamma\propto \gamma$, with $\gamma \ll 1$. Starting from the strongly disordered limit, Fig. 7 in Ref.~\cite{Markos-Muttalib-Wolfle05} suggests a sharp sigmoidal increase in $\gamma$ as disorder is decreased; this signals a transition from the strongly disordered insulating regime towards a weakly disordered metallic regime. Finally
in the metallic regime corresponding to $\gamma \sim 1$, the parameter $\Gamma$ is expected to be very large, although there is no numerical guideline on its $\gamma$-dependence. 
A simple one-parameter model that incorporates the strongly disordered insulating limit as well as the rapid change at the transition as suggested by the numerical studies is given by $\Gamma= a\gamma/[1+\ln\frac{1-\gamma}{\gamma}]$, where $a$ is a phenomenological parameter that loosely characterizes the transition point. 
In the spirit of a toy model, we do not try to fix $a$.  Instead, since our numerical results converge progressively slowly for $\gamma \le 0.5$, we choose $a=0.01$ which generates a transition for $\gamma \sim 0.73$. Starting from the insulating side and systematically increasing $\gamma$, we stop where $\Gamma$ diverges [at $\gamma = e/(1+e)$], and therefore reaches the metallic limit. Note that it is easy to construct a model with more parameters to include the weakly disordered (metallic) regime within this formulation, but since our focus is near the transition, which occurs at strong disorder, we will use the simplest one-parameter model discussed above.

The effect of the logarithm in $V(x,\gamma)$ is two-fold: First, it provides a starting non-monotonicity when combined with the dominant linear single-particle potential. Second, it removes any divergence at the origin.  Thus unlike the CR model, a metallic regime in this case will correspond to a peak in the density near the origin (instead of a diverging density), while an insulating regime will correspond to zero or exponentially small density (a gap) over a finite range near the origin. The metal-to-insulator transition in this case will therefore correspond to the destruction of the peak in the density of eigenvalues near the origin. 

Since there is no divergence at the origin, we use the soft-edge formalism and compute the eigenvalue densities for different values of $\gamma$. Note that in this phenomenological model, both the two-particle interaction term and the single-particle potential change as $\gamma$ is changed. 
Figure  \ref{density_MMW_sigmoid} shows the change in the density as $\gamma$ is increased systematically. At $\gamma=0.7$ the density has a large gap near the origin and is spread out with no peak. As $\gamma$ increases,  the gap becomes smaller and the density starts to develop a peak near the origin. The peak becomes very large at $\gamma=0.73105$, which is the largest value our model allows us to consider. Thus there is a clear ``transition" in the density from zero near the origin to a large peak.
 
\begin{figure}
\begin{center}
\includegraphics[width=0.4\textwidth]{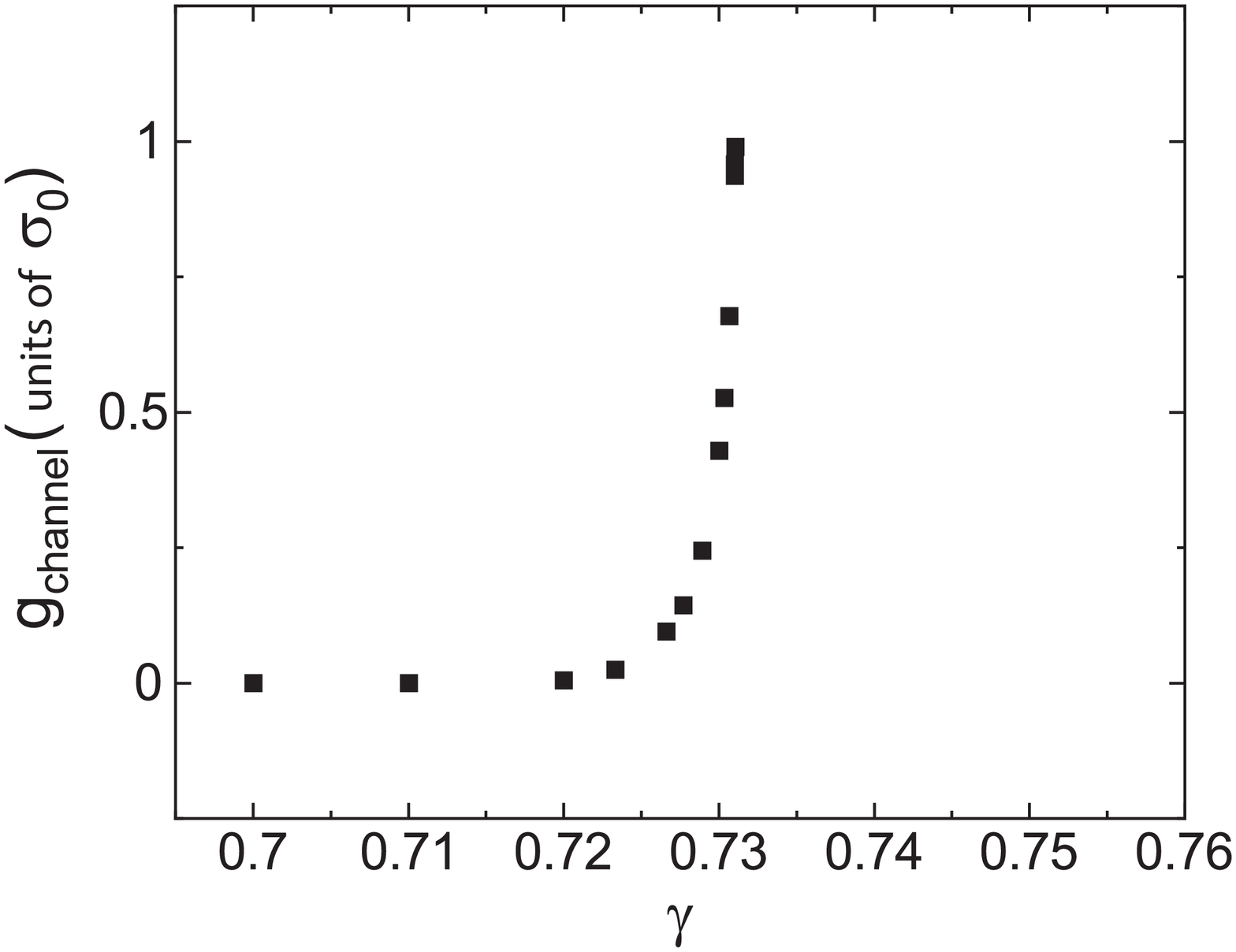}
\end{center}
\caption{
(Color online)
The average conductance $g_{channel}$, computed from eigenvalue densities for different $\gamma$ from Figure \ref{density_MMW_sigmoid}.}
\label{conductance_MMW_sigmoid}
\end{figure}

Clearly, our simplified solvable toy-models can not provide a quantitative description of a three-dimensional disordered system. Nevertheless, the toy model discussed here can provide qualitatively correct behavior for some of the quantities that are not sensitive to the details of the system parameters. Here we use Eq. (\ref{g}) to compute $g_{channel}$, the average conductance per channel (in units of the conductance quantum $e^2/\hbar$). 
Figure \ref{conductance_MMW_sigmoid}  shows how this quantity changes with $\gamma$. 
At $\gamma=0.70$ where the density has a large gap near the origin, the conductance is very small, and it remains small as long as the gap remains appreciable, up to $\gamma=0.72$.  Beyond 
$\gamma = 0.725$ the gap in the density starts to close and a peak near the origin starts to grow, and the conductance starts to increase rapidly.  
It reaches the value $g_{channel} \sim 1$ for $\gamma=0.73105$ which corresponds to the metallic regime. Thus a transition in the density from a peak near the origin to a large gap  can be associated with a metal-to-insulator transition in the conductance.


\section{Summary and conclusion} \label{sec:6}

The eigenvalue density of $\gamma$-ensembles has previously been computed by solving the corresponding Riemann-Hilbert problem. 
In this paper we use the same method to explore the role of the parameter $\gamma$  by considering various solvable  toy models. First, we show that
for different values of $\theta$ between $1$ and $2$, the effective potentials for linear as well as quadratic single-particle potentials can become non-monotonic near the origin for  $\gamma < 1$. The minimum of the effective potential shifts further away from the origin as $\gamma$ is decreased systematically. Second, we show that in a CR type model with a fixed non-monotonicity, reducing $\gamma$ can give rise to a transition from a diverging to a non-diverging density. Finally, we show that a toy model that includes a linear as well as a logarithmic single-particle potential as suggested for three-dimensional disordered conductors, $\gamma\sim 1$ gives conductance $g_{channel}\sim 1$, while $\gamma \ll 1$ corresponds to $g_{channel} \ll 1$. For our particular choice of the model, it also shows a rapid change in the conductance at the transition region between the two limits. While this by itself cannot describe a true metal to insulator transition, it provides a framework where in principle one should be able to study the full distribution of conductances $P(g)$ across a metal-insulator transition. 
This is because given a jpd $p(\{x_a\})$  of the eigenvalues, the distribution of conductances $P(g)$ can be expressed as \cite{Markos-Muttalib-Woelfle-Klauder04} 
\be
P(g)=\int \; \prod_a^N dx_a p(\{x_a\})\delta \left(g-\sum_a \frac{1}{\cosh^2 \sqrt{x_a}}\right).
\ee    
Considering the transition in terms of the full distribution rather than in terms of the average (or typical) conductance is particularly important. This is because even in quasi one-dimension, where $\gamma=1$ for all disorder \cite{Beenakker97} and therefore no transition exists \cite{Abrahams79}, $P(g)$ has a highly asymmetric ``one-sided log-normal distribution" at the crossover point \cite{Muttalib-Woelfle99}, which is expected to remain qualitatively valid in three dimensions near the metal-insulator transition that happens at a critical value $\gamma=\gamma_c < 1$. It is also known from numerical studies in three dimensions that at strong disorder, $P(g)$ has a large variance as well as a finite skewness \cite{Douglas-Muttalib09}. The solvable $\gamma$-ensembles with appropriate single-particle potentials provide a possible framework to analytically study a broad and highly asymmetric distribution of conductances across a transition.

As a by-product, we find that the limit $\theta \rightarrow 1$ also corresponds to the Laguerre $\beta$-ensembles.  This allows us to use the model to numerically compute the eigenvalue density for Laguerre $\beta$-ensembles for all $\beta > 1$.  The results agree with various expected analytical expressions including the ones from the exact analytical solution to the RH problem for $\theta=1$. This shows the applicability of our method for general $\gamma$-ensembles with different values of $\theta >1$ and $\gamma > 0$.            

\section{Acknowledgment}
DW acknowledges support from Ministry of Education, Singapore AcRF Tier 1 Grant No. R-146-000-262-114 and National Natural Science Foundation of China (NSFC) Grant No. 11871425.

\appendix


\section{Laguerre $\beta$-ensembles}

The Laguerre $\beta$-ensembles are characterized by the jpd 
\begin{equation}
\begin{aligned}
  &p(\{x_i\}) \propto \prod_{i=1}^Nw(x_i)\prod_{i<j}|x_i-x_j|^{\beta},\\
  &w(x) = e^{-N \frac{\beta}{2}x}, \quad \beta > 1.
\label{Laguerre_beta_jpd}
\end{aligned}
\end{equation}
The limiting eigenvalue density of Laguerre $\beta$-ensembles for $\beta =1,2,$ and $4$ is known analytically \cite{Forrester94,Forrester93}, and 
later it was shown \cite{Baker-Forrester97, Johansson98} that the same expression is also valid for all values of $\beta$. In Eq. (\ref{gamma_ensemble_jpd}) if we take limit $\theta \rightarrow 1$ and $V(x)=\frac{\beta}{2}x$, we get the jpd of Laguerre $\beta$ ensembles with $\beta = 1 + \gamma$. Thus  in the analysis of Sec. \ref{sec:3}, if we take $\theta \rightarrow 1$ and $V(x)=\frac{\beta}{2}x$, we can compute the eigenvalue density for Laguerre $\beta$ ensembles for any $\beta > 1$. Note that Eqs. (\ref{f_integral_eqn})--(\ref{density_hard_edge}) are valid only for $\theta>1$. By choosing $\theta = 1.0001$ for the $\theta \rightarrow 1$ limit, we can obtain numerical results valid for the $\beta$-ensembles. Later in this appendix we analytically solve the RH problem explicitly for the $\theta=1$ case and show that the results are consistent with numerical solution for $\theta \rightarrow 1$. As $\theta \rightarrow 1$ the shape of contour $\nu$ approaches a circle. 

Once the contour and the mapping (and consequently the inverse mapping) is known, we solve Eq. (\ref{c_hard_edge}) and Eq. (\ref{f_integral_eqn}) self-consistently to find $f(x;\beta)$. Then the effective potential and the eigenvalue density are computed with Eq. (\ref{V-effective}) and Eq. (\ref{density_hard_edge}), respectively, for $\beta = 1 + \gamma$.

\subsection{Eigenvalue density for Laguerre $\beta$ ensembles}

\begin{figure}
\begin{center}
\includegraphics[width=0.4\textwidth]{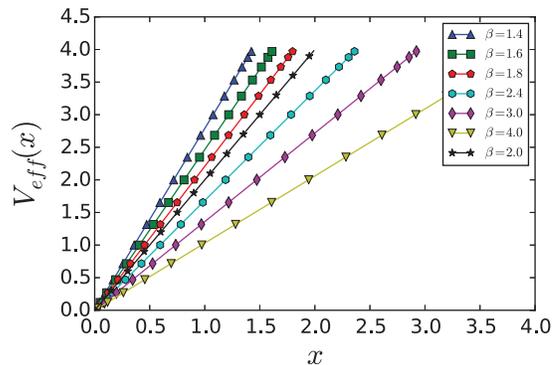}
\end{center}
\caption{
(Color online) 
Effective potential for different $\beta$ and $V(x)=2x$.}
\label{potential}
\end{figure}

\begin{figure}
\begin{center}
\includegraphics[width=0.4\textwidth]{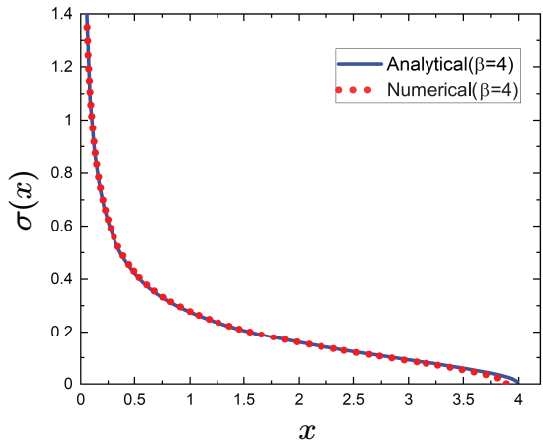}
\end{center}
\caption{
(Color online) 
density for $\beta=4$. The dotted line shows numerical result (see Eq. (\ref{beta_density_scaled})) compared to analytical result shown with bold line.}
\label{density_comparison}
\end{figure}

\begin{figure}
\begin{center}
\includegraphics[width=0.4\textwidth]{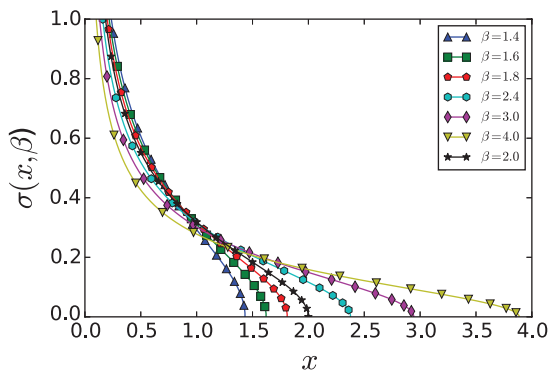}
\end{center}
\caption{
(Color online) 
Densities for different $\beta$ and $V(x)=2x$.}
\label{density}
\end{figure}

In Figure \ref{potential} we show the effective potentials for different $\beta$ over the full support of density. The effective potential becomes less and less converging as $\beta$ increases from $1.4$ to $4$ (or $\gamma$ increases from $0.4$ to $3$). Figure \ref{density} shows the densities calculated from equation (\ref{density_hard_edge}) for different values of $\beta$. These numerical results also agree very well (see Fig. \ref{density_comparison}) with the analytical expression 
\begin{equation}
\sigma(x)=
\begin{cases}
  \frac{2}{\pi}\frac{1}{\beta}(\frac{\beta-x}{x})^{\frac{1}{2}}, & \text{for } 0<x<\beta, \\
  0, & \text{for } x\geq\beta,
\end{cases}
\label{beta_density_scaled}    
\end{equation}
with the density diverging near the origin as $\sigma(x) \to \frac{2}{\pi}\beta^{-\frac{1}{2}}x^{-\frac{1}{2}}$ as  $x \rightarrow 0$.

\begin{figure}
\begin{center}
\includegraphics[width=0.4\textwidth]{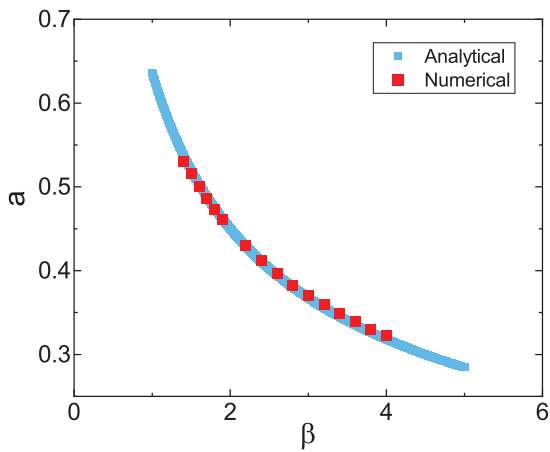}
\end{center}
\caption{
(Color online)
Densities near the origin are fitted to function $\sigma(x;\beta)=ax^b$. The points show prefactors $a$ for different $\beta$. 
The solid line shows the analytical result from Eq. (\ref{beta_density_scaled}).}
\label{prefactor}
\end{figure}

Figure \ref{density} shows that the support of the densities increase as $\beta$ increases. The numerical densities near origin when fitted to curve $\sigma(x;\beta)=ax^b$ show that the exponents b are all $-\frac{1}{2}$ for different $\beta$. Figure \ref{prefactor} shows the prefactors $a$ as function of $\beta$.     


\subsection{RH problem for $\theta = 1$} 

In this section we derive the analytic form of the effective potential for the Laguerre $\beta$ ensemble by exactly solving the RH problem for $\theta =1$, $\gamma >0$. The external potential for the Laguerre $\beta$ ensemble is $V(x)=\frac{\beta}{2}x = \frac{1+\gamma}{2}x$. For $\theta =1$, contour $\nu$ is a unit circle in the complex plane centered at origin. The regions inside and outside the contour $\nu$ are both mapped onto the same complex region $\mathbb{C}\backslash [0,b]$,  similar to the contour shown in Fig. \ref{mapping_schematic}. Every point on the contour is mapped onto a point on the real line in $[0,b]$.

When $\theta =1$, Eq. (\ref{complex_transforms}) gives $g(z)=\tilde{g}(z)$. $M(s)$ is then defined as
\begin{equation}
M(s)\equiv
\begin{cases}
  G[J_c(s)], & \text{for } s\in\mathbb{C}\backslash \bar{D}, \\
  G[J_c(s)], & \text{for } s\in D\backslash[-1,0].
\end{cases}
\label{M_def_theta_1}    
\end{equation}
Now since $\tilde{g}_+(x)=g_+(x)$,  region (1) and region (3) in the schematic mapping are one and the same. Similarly $\tilde{g}_-(x)=g_-(x)$ means region (2) and region (4) are the same. In terms of functions $M(s)$ these relations can be written as $M_+(s_1)=M_+(s_2)$ and $M_-(s_1)=M_-(s_2)$ (see Fig. \ref{mapping_schematic}).  With $N(s)\equiv M(s)J_c(s)$, Eq. (\ref{Npm}) now becomes
\begin{equation}
    \begin{split}
      (1+\gamma)[N_+(s_1)+ N_-(s_1)] 
      = {}& 2V^{\prime}[J_c(s)]J_c(s), \cr  
    \end{split}
    \label{Npm_theta_1}
\end{equation}
where $J_c(s) = J_c(s_1) = J_c(s_2)=x \in [0,b]$. With $f[J_c(s)]$ defined according to Eq. (\ref{eq:defn_f}) and $V(x)=\frac{\beta}{2}x = \frac{1+\gamma}{2}x$ for Laguerre $\beta$ ensembles, we finally get,
\begin{equation}
\label{V_eff_theta_1}
  f(x)= x, \;\;\; V_{\eff}(x)= x.
\end{equation}
 Equation (\ref{V_eff_theta_1}) tells us that the non-monotonicity of the effective potentials previously shown for $\gamma<1$ should disappear when $\theta = 1$. 

In the RH problem for $\theta =1$, if we choose $V(x)=2x$ instead of $\frac{\beta}{2}x$, Eq. (\ref{Npm_theta_1}) gives $f(x)=V_{\eff}(x)=\frac{4}{1+\gamma}x$. The numerical results obtained for the effective potential of the $\gamma$ ensemble with $\theta=1.0001$ and $V(x)=2x$ agree very well with this analytic expression (see Fig. \ref{potential}).




\section{soft-edge formalism} 

\begin{figure}
\begin{center}
\includegraphics[width=0.5\textwidth]{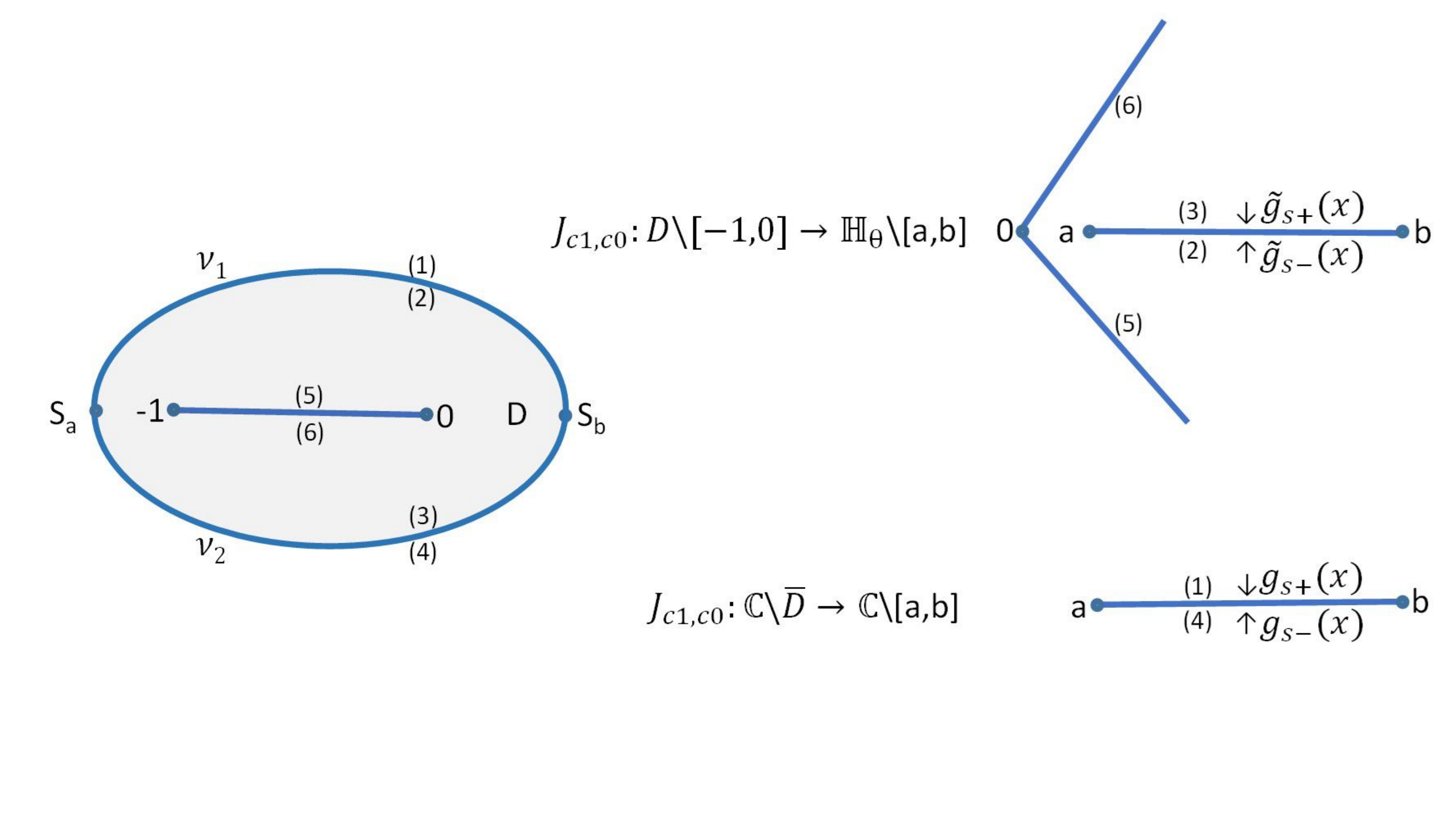}
\end{center}
\caption{
(Color online) 
Schematic figure for the mapping of JT for a soft-edge problem.}
\label{mapping_soft_schematic}
\end{figure}

For soft-edge support, Joukowsky transfromation is given by,

\begin{equation}
\begin{aligned}
  J_{c_1,c_0}(s) = {}& (c_1s+c_0)(\frac{s+1}{s})^{\frac{1}{\theta}}
\end{aligned}
\label{joukowsky_soft}
\end{equation}

where $s$ is a complex variable. Note that the transformation now contains two
parameters $c_0$ and $c_1$ to include the two supports for the soft-edges given by $[a, b]$ where both $a$ and $b$ are real numbers such that $b > a > 0$. The contour $\nu$ (which is a locus of points in the complex plane mapped onto the real line) corresponding to $J_{c_1,c_0}(s)$ is given by,

\bea
\label{contour_soft-edge}
x(\phi)=\frac{r(\phi)\cos\phi -1}{r^2(\phi)-2r(\phi)\cos\phi +1}, \cr
y(\phi)=\frac{-r(\phi)\sin\phi -1}{r^2(\phi)-2r(\phi)\cos\phi +1}
\eea
and $r(\phi)$ solves
\bea
r^2(\phi) &+&\bigg[\frac{c_1}{c_0}\cos\phi-\frac{\sin\phi}{\tan\frac{\phi}{\theta}}-2\cos\phi \bigg]r(\phi)\cr
&+& 1-\frac{c_1}{c_0} =0. 
\eea

where $0<\phi<2\pi$ is the argument of $\frac{s+1}{s}$ in the complex plane. Schematic Fig. \ref{mapping_soft_schematic} shows contour $\nu$ and mapping by the JT $J_{c_1,c_0}(s)$.

The complex transforms are now defined on the soft-edge support,
\begin{equation}
\begin{aligned}
  g_s(z) \equiv {}& \int_a^b \log(z-x)d\mu(x), && z \in \mathbb{C}\backslash(-\infty,b], \\
  \tilde{g_s}(z) \equiv {}& \int_a^b \log(z^{\theta}-x^{\theta})d\mu(x), && z \in \mathbb{H}_\theta\backslash(a,b],
\end{aligned}
\label{complex_transforms_soft}
\end{equation}
with their derivatives $G_s(s)\equiv {g_s}^{\prime}(s)$, $\tilde{G_s}(s)\equiv \tilde{g_s}^{\prime}(s)$ and the function $M_s(s)$ as,
\begin{equation}
M_s(s)\equiv
\begin{cases}
  G_s[J_{c_1,c_0}(s)] & \text{for } s\in\mathbb{C}\backslash \bar{D} \\
  \tilde{G_s}[J_{c_1,c_0}(s)] & \text{for } s\in D\backslash[-1,0];
\end{cases}
\label{M_def_soft}    
\end{equation}
the sum and difference of the EL equations can be written as
\begin{equation}
\begin{split}
  {}& M_{s+}(s_1)+\gamma M_{s-}(s_1)+M_{s-}(s_2)+ \gamma M_{s+}(s_2) \\
  {}& = 2V^{\prime}[J_{c_1,c_0}(s)], \\
  {}& M_{s+}(s_1)-M_{s-}(s_2)+M_{s-}(s_1)-M_{s+}(s_2) 
  = 0.
\end{split}
\label{M_EL_soft}
\end{equation}
Here $s_1 \in \nu_1$ and $s_2 \in \nu_2$ (see Fig. \ref{mapping_soft_schematic}). Equation (\ref{M_EL_soft}), together with some of the limits of $M_s(s)$, form the RH problem for $M_s(s)$. The RH problem in terms of $N_s(s)\equiv M_s(s)J_{c_1,c_0}(s)$ is then as follows.

\vspace{0.2cm}

\paragraph*{RH problem for $N$:}

\begin{itemize}
\item
  $N_s$ is analytic in $\compC \setminus \nu$.
\item 
      $N_{s+}(s_1)+\gamma N_{s-}(s_1)+N_{s-}(s_2)+\gamma N_{s+}(s_2)$\\
      $=  2V^{\prime}[J_{c_1,c_0}(s)]J_{c_1,c_0}(s)$ 
      \begin{equation}
      N_{s+}(s_1)-N_{s-}(s_2)+N_{s-}(s_1)-N_{s+}(s_2) 
      =  0.
    \label{Npm_soft}
  \end{equation}
\item 
  $N_s(0) = \theta$, $N_s(-1)=0$ and $N_s(s) \to 1$ as $s \to \infty$.
\end{itemize}
We further define a function $f_s$ such that
\begin{equation} \label{eq:defn_f_soft}
f_s[J_{c_1,c_0}(s)]\equiv N_{s+}(s)+N_{s-}(s).
\end{equation}
This gives the solution to the RH problem of $N_s(s)$ as
\begin{equation}
N_s(s)=
\begin{cases}
  \frac{-1}{2\pi i}\oint_{\nu}\frac{f_s[J_{c_1,c_0}(\xi)]}{\xi -s}\; d\xi +1, & s\in \mathbb{C}\backslash \bar{D} \cr
\frac{1}{2\pi i}\oint_{\nu}\frac{f_s[J_{c_1,c_0}(\xi)]}{\xi -s}\; d\xi -1, & s\in D\backslash [-1,0].
\end{cases}
\label{N_def_contr_soft}
\end{equation}
Also from the RH problem for $N_s(s)$, the constants $c_1$ and $c_0$ of the JT in Eq. (\ref{joukowsky_soft})  satisfy the equations
\begin{equation}
\label{c1_c_0_soft_edge}
\begin{split}
\frac{1}{2\pi i}{\displaystyle \oint_{\nu}^{}}\frac{f_s[J_{c_1,c_0}(s)]}{s}ds= {}& 1+\theta, \\ 
\frac{1}{2\pi i}{\displaystyle \oint_{\nu}^{}}\frac{f_s[J_{c_1,c_0}(s)]}{s+1}ds= {}& 1
\end{split}
\end{equation}
Thus the sum equation in the RH problem for $N_s(s)$ can be rewritten as,
\begin{equation}
\begin{split}
{}&(1-\gamma)(N_{s+}(s_1)+N_{s-}(s_2))+2\gamma f_s[J_{c_1,c_0}(s)] \\ 
{}&=2V^{\prime}[J_{c_1,c_0}(s)]J_{c_1,c_0}(s).
\end{split} 
\end{equation}
Defining the inverse mapping of JT as,
\begin{equation}
s=J_{c_1,c_0}^{-1}(x)=h_s(x).
\label{inversemap_soft}
\end{equation}
with $(s_1)_+ = h_s(y) \ ; \ (s_2)_- = \bar{h_s}(y) \ ; \ s_1 = h_s(x) \ \text{and} \ s_2 = \bar{h_s}(x) $, we substitute for $[N_{s+}(s_1)+N_{s-}(s_2)]$ using Eq. (\ref{N_def_contr_soft}) and the inverse mapping. We finally get the integral equation,
\begin{equation}
\label{f_integral_eqn_soft}
  f_s(y;\gamma) = \frac{V^{\prime}(y)y}{\gamma} 
  -\frac{1-\gamma}{\gamma}\bigg[1 + \frac{1}{2\pi}\int_a^b f_s(x;\gamma)\phi_s(x,y)dx \bigg],
\end{equation}
where
\begin{equation}
  \phi_s(x,y) = \Im\bigg[ \left( \frac{1}{h_s(y) - \overline{h_s}(x)} + \frac{1}{\overline{h_s}(y) - \overline{h_s}(x)} \right) \overline{h_s}^{\prime}(x) \bigg].
\end{equation} 
We solve Eq. (\ref{f_integral_eqn_soft}) for $f_s(y;\gamma)$ and Eq. (\ref{c1_c_0_soft_edge}) for $c_1$ and $c_0$ numerically, self-consistently. The new effective potential $V_{\eff}(x;\gamma)$ is related to $f_s(x;\gamma)$ by
\begin{equation}
V'_{\eff}(x;\gamma)=\frac{f_s(x;\gamma)}{x}.
\label{V-effective_soft}
\end{equation}
 The eigenvalue density for this effective potential is given by \cite{Alam-Muttalib-Wang-Yadav20},  
\begin{equation}
\label{density_soft_edge}
\begin{split}
  \sigma_s(y;\gamma)= {}& \frac{-1}{2{\pi}^2 \gamma y}\int_{b}^a xV'_{\eff}(x;\gamma)\chi_s(x,y) dx,\\
  \chi_s(x,y)= {}& \Re\bigg[ \bigg( \frac{1}{\overline{h_s}(y) - h_s(x)}-\frac{1}{h_s(y) - h_s(x)} \bigg){h_s}^{\prime}(x)\bigg].
\end{split}
\end{equation}




\section{Transport in dimensions other than  $d=3$.} 

In the present  work we have focused on transport in three dimensions ($d=3$) only. In the absence of electron-electron interactions, there exists metal-insulator transition only for $d >2$ with $d=2$ being a critical dimension \cite{Abrahams79}. How does the dimensionality enter in our formulation? The short answer is that  the parameter $\gamma$ has a highly non-trivial dimensionality dependence, which results in having, e.g., a true metal-insulator transition in 3D but only a crossover in quasi one-dimension (Q1D). In this appendix, we briefly outline how this dimensionality dependence of $\gamma$ is included in the present formulation. 

When a disordered conductor of cross section $L^{d-1}$ and length $L_z$ is connected to two perfect leads, the scattering states at the Fermi energy defines $N\propto L^{d-1}$ channels. Transport properties are then characterized by the $2N\times 2N$ transfer matrix $M$ that connects the outgoing flux to the incoming flux across $L_z$. Flux conservation and time-reversal symmetry allows one to write $M$ in the general form \cite{Mello-Pereyra-Kumar88}
\bea
M=\left(\begin{array}{cc}u & 0\cr 0& u^* \end{array}\right) \left(\begin{array}{cc}\sqrt{1+\lambda} & \sqrt{\lambda} \cr  \sqrt{\lambda} & \sqrt{1+\lambda} \end{array}\right)
\left(\begin{array}{cc}v & 0\cr 0& v^* \end{array}\right),
\eea
where $u$, $v$ are $N\times N$ unitary matrices and $\lambda$ is a diagonal matrix with non-negative elements. It turns out that in terms of the parameters of $M$, the $N\times N$ matrix $tt^{\dag}$, where $t$ is the  $N$-channel transmission matrix, can be written as \cite{Markos-Muttalib-Wolfle05} 
\be
tt^{\dag}=v^*(1+\lambda)^{-1}v.
\ee
Diagonalizing  $tt^{\dag}$ gives us  $\lambda$ as well as all elements of the $N\times N$ matrix $v$.  The conductance $g$ is then given by $g={\rm Tr}(tt^{\dag})=\sum_{a=1}^N 1/(1+\lambda_a)$, while the  matrix $v$ contains information about the dimensionality of the system in the following way \cite{Markos-Muttalib-Wolfle05} :

Consider the $N\times N$ matrix
\be
\gamma_{ab}=\frac{2K_{ab}}{K_{aa}}; \;\;\; K_{ab}= \langle \sum_{\alpha=1}^{N} |v_{a\alpha}|^2|v_{b\alpha}|^2\rangle ,
\ee
where the angular bracket represents an ensemble average.  This matrix appears in the generalized DMPK (DorokhovMello-Pereyra-Kumar) equation \cite{Markos-Muttalib-Wolfle05, Gopar-Muttalib02, Douglas-Markos-Muttalib14}, 
\bea
\frac{\partial p(\{\lambda\})}{\partial (L_z/l)} &=& \sum_{a=1}^NK_{aa}\frac{\partial}{\partial\lambda_a} \lambda_a(1+\lambda_a) \cr
&\times&  \left[\frac{\partial}{\partial\lambda_a}-\sum_{b\ne a}\frac{\gamma_{ab}}{\lambda_a-\lambda_b}\right]p(\{\lambda\})
\label{generalized_DMPK}
\eea
with $l$ being the mean free path, whose solution gives the evolution of the jpd $p(\{\lambda_a\})$ of the eigenvalues $\lambda$ with length, and where the length and  disorder dependence of the matrix $\gamma_{ab}$ contain information about dimensionality.  

For example in Q1D, for a wire of cross-section  $L\times L$ and length $L_z$ with $L \ll L_z$, the localization length $\xi$ is much larger than the transverse length, $\xi\gg L$, and all channels of transport become equivalent (this is the definition of a Q1D system). This implies that the eigenvectors of $v$ are isotropic. In this case \cite{Markos-Muttalib-Wolfle05}, 
\be
K_{ab}^{Q1D}=\frac{1+\delta_{ab}}{N+1}, \;\;\; \gamma_{ab}^{Q1D}=1.
\ee
Note that $\gamma^{Q1D}=1$ is true for all disorder. 
This is the reason why a jpd of the form Eq. (\ref{DMPK_3D_MMW}) with $\gamma=1$ implies Q1D. The metal to insulator crossover in this case is rather unphysical, it occurs with increasing $L_z$;  $L_z\ll \xi$ being a metal while $L_z\gg \xi$ is an insulator \cite{Beenakker97}. (Note that our formulation is based on a large $N$ limit of the jpd, so it is not valid for strictly $d=1$.)

In higher dimensions, in the absence of isotropy, the matrix $\gamma_{ab}$ is much more complicated and can not be obtained analytically. However, within a tight-binding Anderson model with random site energies (with strength $W$) and nearest neighbor hopping elements, which shows the Anderson metal-insulator transition at a critical disorder $W_c$, it is possible to study the matrix $K_{ab}$ in the space representation (as opposed to a channel representation) numerically as a function of both length and disorder. 
This was done in detail explicitly for $d=3$ in Ref.~[\onlinecite{Markos-Muttalib-Wolfle05}].  The results show that to a good approximation, one can consider only the most dominant element $\gamma_{12}=\gamma^{3D}$. Indeed, this one parameter generalization of DMPK equation was originally conjectured to be applicable beyond the Q1D regime \cite{Klauder-Muttalib99}. With this approximation for $\gamma_{12}$, the solution to the generalized DMPK equation Eq. (\ref{generalized_DMPK}) gives the jpd for 3D systems as shown in Eq. (\ref{DMPK_3D_MMW}) where $\gamma=\gamma^{3D}$. A finite-size scaling analysis shows that for large $L=L_z$, as disorder is increased, $\gamma^{3D} \lesssim 1$ for $W < W_c$, and $\gamma^{3D} \ll 1$ for $W > W_c$, dropping sharply near $W=W_c$ (see, e.g., Figs. 7 and 21 in Ref.~[\onlinecite{Markos-Muttalib-Wolfle05}]).  The phenomenological model considered in Sec. V uses these results to relate the parameter $\Gamma(\gamma)$ to disorder in 3D. 

The matrix $K_{ab}$ has not been studied in any other dimensions. However,
given the fact that in 2D in the limit $L=L_z \to\infty$ the system is always an insulator for all finite disorder, one can conjecture that $\gamma^{2D}$ remains much smaller than 1 for all disorder, never reaching the metallic limit. If the explicit disorder dependence of $\gamma^{2D}$ is known, our formulation can be adapted to study the conductance distribution $P(g)$ in 2D. Similarly, given that metal-insulator transitions exist  for all higher dimensions, we expect that the disorder dependence of $\gamma^{dD}$ for $d>3$ will be qualitatively similar to $\gamma^{3D}$, but with different (dimension dependent) critical disorder $W_c$. 

In summary, for an arbitrary $d$-dimensional conductor, the current formulation  can be used to obtain the full distribution of conductances for all disorder if the explicit disorder dependence of the parameter $\gamma^{dD}$ can be obtained from numerical studies of the matrix $K$ using the tight-binding Anderson model in $d$-dimensions.  



\providecommand{\newblock}{}


\begin{thebibliography}{10}
\bibitem{note} For a disordered conductor with $N$ channels, the eigenvalues $X_i$ of the matrix $X=\frac{1}{4}[MM^{\dag}+M^{\dag}M-2I]$ where $M$ is the $2N\times 2N$  transfer matrix characterizing the conductor, the conductance is $g=\sum_i^N\frac{1}{1+X_i}$  \cite{Muttalib-Pichard-Stone87}. The variables $x_i$ used in (\ref{gamma_ensemble_jpd}) is related to $X_i$ by $X_i=\sinh^2\sqrt{x_i}$ \cite{Markos-Muttalib-Wolfle05}.
\end{thebibliography}

\end{document}